\documentclass[]{interact}
\usepackage{hyperref}
\usepackage{epstopdf}
\usepackage[caption=false]{subfig}
\graphicspath{{figures/}}

\usepackage[numbers,sort&compress,merge]{natbib}
\bibpunct[, ]{[}{]}{,}{n}{,}{,}
\newcommand{\dist}[1]{f_{#1}(#1)}

\newcommand{\matcj}{\mathrm{ce}_j(\tau)}
\newcommand{\matsj}{\mathrm{se}_j(\tau)}
\newcommand{\matcjp}{\mathrm{\dot{ce}}_j(\tau)}
\newcommand{\matsjp}{\mathrm{\dot{se}}_j(\tau)}
\theoremstyle{plain}

\theoremstyle{definition}

\theoremstyle{remark}

\begin{document}

\articletype{ARTICLE}

\title{The energy distribution of an ion in a radiofrequency trap interacting with a nonuniform neutral buffer gas\footnote{Dedicated to Prof. Timothy P. Softley on the occasion of his 60$^\text{th}$ birthday.} }

\author{
\name{I. Rouse\textsuperscript{a} and S. Willitsch\textsuperscript{a}}
\affil{\textsuperscript{a} Department of Chemistry, University of Basel, Klingelbergstrasse 80, 4056 Basel, Switzerland}
}

\maketitle

\begin{abstract}
An ion in a radiofrequency (rf) trap sympathetically cooled by a simultaneously trapped neutral buffer gas exhibits deviations from thermal statistics caused by collision-induced coupling of the rf field to the ion motion. For a uniform density distribution  of the buffer gas, the energy distribution of the ion can be described by Tsallis statistics. Moreover, runaway heating of the ion occurs if the buffer gas particles are sufficiently heavy relative to the ion. In typical experiments, however, ultracold buffer gases are confined in traps resulting in localised, non-uniform density distributions. Using a superstatistical approach, we develop an analytical model for an ion interacting with a localised buffer gas. We demonstrate theoretically that limiting collisions to the centre of the ion trap enables cooling at far greater mass ratios than achievable using a uniform buffer gas, but that an upper limit to the usable mass ratio exists even in this case. Furthermore, we analytically derive the energy distribution for an ion interacting with a buffer gas held in a harmonic potential. The analytical distribution obtained is found to be in excellent agreement with the results of numerical simulations. 
\end{abstract}

\begin{keywords}
Ion-neutral collisions, hybrid traps, sympathetic cooling, superstatistics, Tsallis statistics.
\end{keywords}

\section{Introduction}
The combination of cooling and trapping techniques has enabled the production of confined ultracold matter: particles localised in space and with mean kinetic energies corresponding to less than a millikelvin with applications ranging from spectroscopy to quantum computing \cite{jin12a}. Hybridising different trapping architectures enables the co-trapping of multiple species, for example, the simultaneous confinement of charged and neutral particles in ion-neutral hybrid traps. Hybrid traps allow for the study of collisions between charged and neutral particles under highly controlled conditions providing insight into the fundamental interactions between them \cite{haerter14a,tomza17a,willitsch15a,willitsch17a}. In these experiments, the ions are typically confined using a radiofrequency (rf) trap, in which an oscillating electric potential enables dynamic confinement, whereas the neutral atoms are confined using magnetic, magneto-optical, or optical dipole traps.  It could be expected that if a single ion is in contact with a neutral buffer gas, then due to the exchange of energy between the ion and the buffer gas through elastic collisions the ion should reach thermal equilibrium with the buffer gas. That is, the ion is sympathetically cooled by the buffer gas particles, and it would be expected that the energy distribution of the ion is given by a thermal distribution of the form,
\begin{equation} \label{eq:energyDistConditional}
f_E(E|T_b) = \frac{E^k}{ (k_B T_b)^{k+1}  \Gamma(k+1) } e^{- \frac{E}{k_B T_b}}, 
\end{equation}
where $T_b$ is the temperature of the buffer gas, $m_i$ is the mass of the ion, $E^k$ represents the density of states, and the notation $E|T_b$ indicates that this is the energy distribution for a specific value of the temperature. It has been previously shown that this does not occur because of the transfer of energy from the motion of the ion driven by the rf potential (micromotion) to its secular (thermal) motion as a result of collisions \cite{moriwaki92a,devoe09a,zipkes11a,chen14a}. This process of micromotion interruption leads to fluctuations in the energy of the ion sufficient to drive the system from thermal equilibrium, and consequently a non-thermal energy distribution is obtained. If the buffer gas is uniformly distributed in space, it has been observed that the resulting distribution follows Tsallis statistics \cite{devoe09a,meir16b,hoeltkemeier16a},
\begin{equation} \label{eq:tsallisEnergyDistFirst}
f_{E}^{(T)}(E) = \left(\frac{n_T}{\langle\beta\rangle}\right)^{-k-1}  \frac{\Gamma (k+n_T+1)}{\Gamma (k+1) \Gamma (n_T)}     \frac{E^k}{ \left(\frac{\langle\beta\rangle E}{n_T}+1\right)^{k + n_T + 1}} 
\end{equation} 
defined in terms of a scale parameter $\langle \beta \rangle$ and the Tsallis exponent $n_T$. When $E$ is large, the Tsallis distribution asymptotically approaches a power-law of the form $f_E(E) \propto E^{-(n_T + 1)}$, in contrast to the exponential decay of a standard thermal distribution. Using the formalism of superstatistics \cite{beck01a,beck03a}, we have analytically shown that Tsallis statistics are a suitable model to describe the energy distribution for an ion interacting with a uniform buffer gas  \cite{rouse17a,rouse18a}. 

Models of sympathetic cooling of ions by a buffer gas of uniform density predict that, if the ratio of the mass of the buffer-gas particles to the mass of the ion, $\tilde{m} = m_b  / m_i$, is above a  value of approximately $1.2$, then the ion will on average gain energy from each collision and so cannot be effectively cooled, corresponding to a Tsallis distribution with $0 < n_T < 1$ \cite{moriwaki92a,devoe09a,chen14a}. This is true if collisions occur with equal probability at all points along the trajectory of the ion. However, it has been demonstrated through numerical simulations as well as in experiments that this limit can be overcome by ensuring that collisions are more likely when the ion is close to the centre of the trap \cite{goodman12a,hoeltkemeier16a,dutta17a}. At this point, the secular velocity is large compared to the velocity of the micromotion, and so collisions on average cause a greater reduction in the energy of the ion than if they occur further away from the centre. By ensuring that collisions preferentially occur close to the trap centre, the cooling remains efficient at much greater mass ratios \cite{dutta17a,hoeltkemeier16a,hoeltkemeier16b}.  Moreover, under these conditions, the energy distribution has been found to deviate from Tsallis statistics \cite{hoeltkemeier16a,hoeltkemeier16b}, and has been modelled using a Tsallis distribution multiplied by an exponential function,
\begin{equation} \label{eq:et}
f_E^{(ET)}(E) \propto f_{E}^{(T)}(E)  e^{-E/E_a} ,
\end{equation}
where $E_a$ is a characteristic energy scale. We refer to Eq. (\ref{eq:et}) as the exponential-Tsallis (ET) distribution, as when $E < E_a$ the distribution resembles Tsallis statistics, but for $E > E_a$ it exhibits an exponential decay instead of the power-law tail of standard Tsallis statistics. The exponential-Tsallis distribution is a special case of the compound confluent hypergeometric distribution as discussed in Ref.~\cite{nadarajah07a}, but the use in Ref.~\cite{hoeltkemeier16a} appears motivated as an empirical modification of the standard Tsallis distribution.

Based on our previous work \cite{rouse17a,rouse18a}, we develop here an analytical model to characterise in detail the effects of non-uniform buffer-gas-density distributions on the outcome of ion-neutral collisions. We confirm that limiting collisions to the centre of the trap results in efficient cooling at a much greater range of mass ratios, but that for a sufficiently heavy buffer gas the ion may still gain energy on average in each collision. By employing the method of superstatistics \cite{beck01a,beck03a}, we analytically derive the energy distribution of an ion undergoing elastic collisions with a localised buffer gas in the limit in which the trapping frequencies of the buffer gas are much smaller than the trapping frequencies of the ion. We find analytically that it resembles Tsallis statistics at low energy and asymptotically approaches an exponential decay of the form $e^{-\sqrt{E}}$ for large values of the energy, similar but not identical to the asymptotic behaviour proposed in Ref.~\cite{hoeltkemeier16a}. This form of the distribution is confirmed using numerical simulations finding an excellent agreement over a wide range of mass ratios. 

\section{Theory}
\subsection{Ion motion}
We assume that the ion is trapped in an ideal harmonic rf trap described by the homogenous Mathieu equation \cite{major05a},
\begin{equation} \label{eq:mathieuHomogenous}
\ddot r_j(\tau)  + [a_j - 2q_j \cos(2 \tau) ] r_j(\tau)=0,
\end{equation}
where $a_j,q_j$ are the Mathieu stability parameters calculated from the trapping potential and $\tau = \Omega t/2$ is the time $t$ scaled by $1/2$ of the rf frequency $\Omega$. The solution is given by \cite{boyce17a}, 
\begin{equation}\label{eq:mathieuPositionHomogenous}
r_j(\tau)   =   A_j [ \mathrm{ce_j}(\tau) \cos \phi_j - \mathrm{se_j} (\tau)  \sin \phi_j],
\end{equation}
where $A_j$ is the amplitude of the motion, $\phi_j$ is the phase, and   $\mathrm{ce}_j(\tau)$, $\mathrm{se}_j(\tau)$ are a pair of linearly independent solutions to Eq.~\eqref{eq:mathieuHomogenous}. The cosine elliptic function $\mathrm{ce}_j(\tau)$ is defined in terms of a Fourier series,
\begin{equation} \label{eq:mathieuFunctionFourier}
\mathrm{ce}_j(\tau) = \sum_m c_{2m,j} \cos[(\beta_j + 2 m)\tau]
\end{equation}
where $\beta_j$ is the characteristic exponent, and the Fourier coefficients are found by a recurrence relation \cite{olver2010a}. The sine elliptic function $\mathrm{se}_j(\tau)$ is defined analogously by replacing $\cos$ by $\sin$ in Eq.~\eqref{eq:mathieuFunctionFourier}. It is useful to separate the motion into the secular motion, consisting of the $m = 0$ terms of the Fourier series expansions of $\mathrm{ce}_j(\tau)$ and $\mathrm{se}_j(\tau)$, and the high-frequency micromotion, $m \neq 0$. The secular motion of the ion is given by,
\begin{equation}
\tilde{r_j}(\tau) = A_j c_0 \cos [\beta_j \tau + \phi_j],
\end{equation}
which can be seen to be a harmonic oscillator of amplitude $A_j c_0$ and a frequency (in terms of $t$ rather than $\tau$) of $\omega_j = \frac{1}{2} \beta_j \Omega$.
As a result of the time-dependent potential, the energy of the ion is not a conserved quantity. However, $A_j$ is conserved, and we may define the secular energy as the energy of a harmonic oscillator of amplitude $A_j c_0$ and secular frequency $\omega_j$,
\begin{equation} \label{eq:chapter6EnergyDef}
E_j =  \frac{m_i}{2} \frac{\Omega^2}{4} A_j^2 \beta^2_j c_{0,j}^2 =\frac{m_i}{2}  \omega_j^2 A_j^2 c_{0,j}^2 .
\end{equation}

\subsection{Ion-neutral collisions} \label{section:energyChangeInCollision}
To simplify the problem, it is assumed that collisions are classical and instantaneous following the Langevin model commonly used for ion-neutral collisions \cite{devoe09a,zipkes11a,chen14a}. Under these conditions, the  trajectory of the ion is defined at all times by Eq.~\eqref{eq:mathieuHomogenous}. The trajectory after a collision must therefore have the same general form as Eq.~\eqref{eq:mathieuPositionHomogenous}, but with the constants of integration $A_j,\phi_j$ updated to new values,
\begin{equation}\label{eq:mathieuPositionHomogenousPostCollision}
r'_j(\tau)   = A'_j [ \mathrm{ce_j}(\tau) \cos \phi'_j - \mathrm{se_j} (\tau)  \sin \phi'_j] ,
\end{equation}
where primes indicate post-collision quantities.  We assume that all collisions are elastic and occur within a hard-sphere model so that the post-collision velocities are given by \cite{devoe09a, zipkes11a, chen14a, rouse17a} ,
\begin{equation} \label{eq:collisionAtomVel1}
\mathbf{v'}  =  \frac{1}{1 + \tilde{m}}  \mathbf{v}   +  \frac{\tilde{m}}{1 + \tilde{m}}  \mathbf{v_b}  + \frac{\tilde{m}}{1 + \tilde{m}}   \mathrm{R}  (\mathbf{v} -\mathbf{v_b} ) ,
\end{equation}
where bold-faced variables indicate vectors, e.g., $\mathbf{v} = (v_x,v_y,v_z)^T$. Here, $\mathbf{v_b}$ is the velocity of the colliding buffer-gas particle, $\tilde{m} = m_b/m_i$ is the buffer gas-to-ion mass ratio, and $ \mathrm{R} $ is a rotation matrix determined by the scattering angles.  The method to derive the post-collision energy has been described elsewhere \cite{rouse17a,rouse18a}, and we simply quote the final result that each component of the post-collision secular energy is given by,
\begin{equation} \label{eq:energyComponents}
\begin{split}
E'_j  = \sum_{(k,l) \in (x,y,z) } &\bigg (  \eta_{jkl} \sqrt{E_k} \sqrt{E_l} + a_{1,jkl} \sqrt{E_k} v_{b,l} \\ &+   a_{2,jkl} v_{b,k} v_{b,l}      \bigg) ,
\end{split}
\end{equation}
where the coefficients $\eta_{jkl}$ and $a_{i,jkl}$ describe the transfer of energy between the motion along the three coordinate axes and between the ion and the colliding particle of buffer gas. The coefficients depend on the elements of the random rotation matrix $\mathrm{R}$, the set of phases $\phi_j$, and the time of collision $\tau$. Although these can be found analytically, the resulting expressions are lengthy and so are not reproduced here, see the supplementary material of Ref.~\cite{rouse18a} for details. For the purposes of this work, it suffices to note that the $\eta_{jkl}$ are dependent on $\cos^2 \phi_j$, $\sin^2 \phi_j$ and $\cos \phi_j \sin \phi_j$. This is a result of the fact that the post-collision secular energy is obtained by squaring the elements of Eq.~\eqref{eq:collisionAtomVel1}, which in turn depend on $\sin \phi_j,\cos \phi_j$ from the definition of the velocity as the derivative of Eq.~\eqref{eq:mathieuPositionHomogenous}.  

Eq.~\eqref{eq:energyComponents} represents a three-dimensional, non-linear recurrence relation for the evolution of the components of the secular energy with each collision which is not analytically tractable. Two simplifications can be made to produce a one-dimensional, linear recurrence relation. Firstly, we make a change of variables of the form,
\begin{equation}
E_j  = E  P_j ,
\end{equation}
where $P_x = \cos^2 \phi_\rho \sin^2 \theta_\rho, P_y  = \sin^2 \phi_\rho \sin^2 \theta_\rho, P_z = \cos^2 \theta_\rho$. We find,
\begin{equation}
\eta_{jkl} \sqrt{E_k} \sqrt{E_l} =  E \eta_{jkl} \sqrt{P_k} \sqrt{P_l},
\end{equation}
enabling the factoring of $E$ out of terms proportional to $\eta_{jkl}$. Secondly, due to the fact that the velocity distribution of the buffer gas is isotropic, any terms linear in $v_{b,l}$ average to zero and can be neglected, i.e., those proportional to $a_{1,jkl}$. Summing over $j$ in Eq.~\eqref{eq:energyComponents} and making these approximations, we arrive at a linear, one-dimensional recurrence relation of the form,
\begin{equation} \label{eq:energyRecurrence}
E' = \eta E + \epsilon,
\end{equation}
where $\eta = \sum_{jkl} \eta_{jkl} P_k P_l$ and $\epsilon = \sum_{jkl}  a_{2,jkl} v_{b,k} v_{b,l}$.  The coefficient $\eta$ represents the multiplication of the energy of the ion by a random amount due to micromotion interruption, whereas $\epsilon$ is an additive term representing the increase in the energy of the ion due to the kinetic energy of the buffer gas. We refer to these as the multiplicative and additive noise, respectively. 

This change of the secular energy as a result of a collision is valid regardless of the density distribution of the buffer gas. However, $\eta$ and $\epsilon$ are defined in terms of the random variables describing the circumstances of a collision and the random rotation of the trajectory, that is, the time $\tau$, the set of phases $\phi_j$, the velocity of the buffer gas, and the elements of the random rotation matrix $R$. If the distributions of these random variables change, then so do the distributions of $\eta$ and $\epsilon$, leading to different statistical outcomes of the collision processes. The elements of the random rotation matrix in the Langevin model correspond to a uniform isotropic rotation and the velocity of the buffer gas is assumed to follow Maxwell-Boltzmann statistics. We assume in general that the density of the buffer gas is sufficiently low that collisions are separated by an interval of time which is large compared to the period of the secular motion, such that consecutive collisions occur independently of each other. For a buffer gas of uniform density, the rate of collisions is independent of the time at which the collision occurs $\tau$ and the phase angles $\phi_j$, as collisions are equally likely at all points along the trajectory of the ion. In this case it suffices to assume that $\tau$ and the set of values $\phi_j$ follow uniform distributions. 

However, if the density of the buffer gas is not uniform, then collisions are more likely for certain values of $r_j(\tau)$, and thus at certain combinations of values of $\tau$ and $\phi_j$. These must therefore be described by a joint distribution $f(\tau,\phi_j)$, which may be expressed as \cite{riley10a},
\begin{equation}
    f(\tau,\phi_j)  = f_{\phi_j|\tau}(\phi_j|\tau) f_{\tau}(\tau).
\end{equation}
We assume that after averaging over all values of $\phi_j$, the probability for a collision to occur is equal for all values of $\tau$, such that the marginal distribution $f_\tau(\tau)$ can be taken to be a uniform distribution. By performing numerical simulations (as described in Section~\ref{section:numericalSimulations} with Mathieu parameter $q_j = 0.5$ and buffer gas radial trapping frequency 1000~Hz) to sample the values of $\tau$ at which collisions occur in a non-uniform buffer gas, this is found to be an excellent approximation.  Thus, all that remains is to find the values of $\phi_j$ given that a collision has taken place.

\subsection{Collisions at the centre of the trap} \label{section:centralCollisions}
We first consider the extreme case in which collisions may occur only at the exact centre of the trap. Although this situation is not physically realistic, it provides insight into the outcome and limitations of sympathetic cooling using a non-uniform buffer gas. For the ion to be at the centre of the trap, we require that $r_j(\tau) = 0$ for each $j \in (x,y,z)$. Using Eq.~\eqref{eq:mathieuPositionHomogenous} for the position of the ion we find,
\begin{equation}
A_j \cos \phi_j ~\matcj - A_j \sin \phi_j~ \matsj = 0.
\end{equation}
Solving this equation for $\phi_j$ for non-zero values of $A_j$ produces,
\begin{equation} \label{eq:phijAtCentre1}
\tan \phi_j = \frac{\matcj}{\matsj}.
\end{equation}
This equation has two possible solutions for $\phi_j \in [0,2\pi)$, corresponding to motion in either the positive or negative direction.

\begin{figure}[tb]
\centering
\includegraphics[width=.9\linewidth]{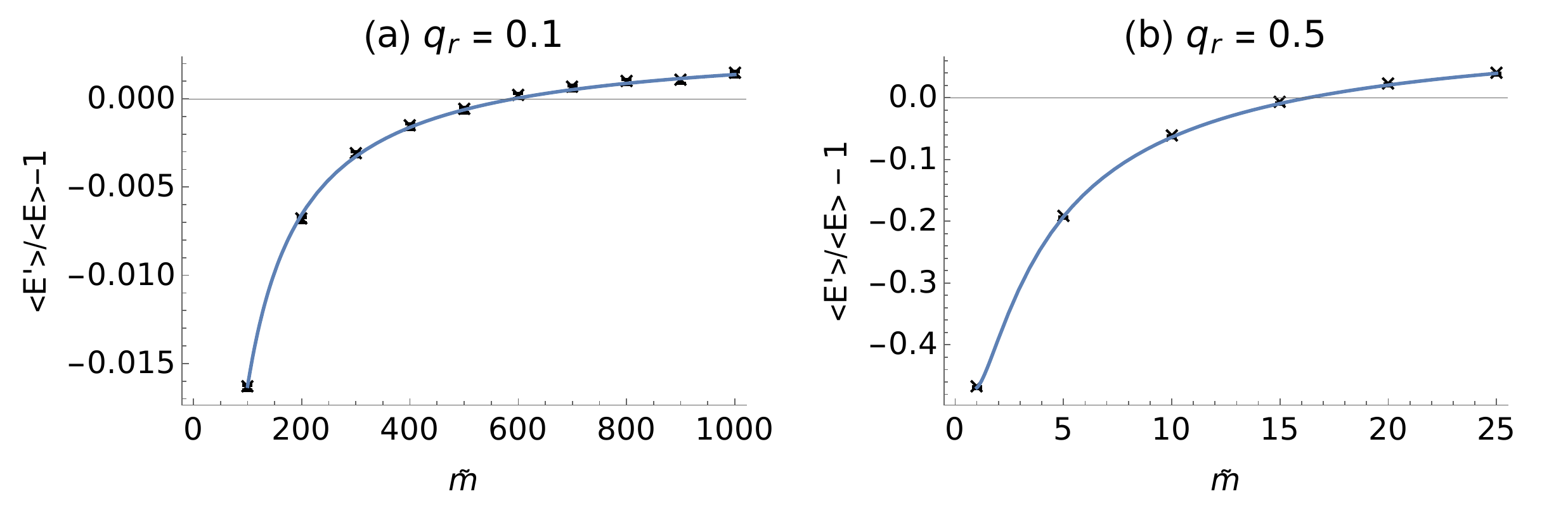}
\caption{The change in the secular energy due to collisions of an ion of mass $m_i$ with a buffer-gas particle of mass $m_b$ as a function of the mass ratio  $\tilde{m} = m_b/m_i$ when the collisions take place at the centre of an rf trap. The trap is taken to be an ideal linear trap with $q_x = -q_y = q_r, q_z = 0, a_x = a_y = -a_z/2, a_z = 0.000625$, and the velocity of the buffer gas prior to the collision is set equal to zero. Two cases are illustrated: (a) $q_r = 0.1$ and (b) $q_r = 0.5$. The solid lines show the analytical expression obtained by averaging over all the collision parameters, while the points give the results obtained from numerical simulations of a collision at the centre of the trap with 100'000 simulations per point. Standard errors are smaller than the plot symbols. In all cases, it has been assumed that the secular energy before the collision is given by a thermal distribution with the same mean energy for the motion along each coordinate axis. Values of $\langle E'\rangle / \langle E \rangle > 1$, i.e., points above the horizontal line, indicate that the ion on average gains energy from the collision due to micromotion interruption. }
\label{fig:energy_ratio_fullLocalisation}
\end{figure}

Using the two allowed values for $\phi_j$ together with the change in energy obtained in the previous section, we derive the ratio of the mean post-collision energy to the mean pre-collision energy for a collision taking place at the exact centre of the trap in Appendix~\ref{section:centralCollisionAppendix}. For a buffer gas of temperature $T_b = 0~$K and under the assumption that the ion is in thermal equilibrium before the collision, i.e., $\langle E_x \rangle = \langle E_y \rangle = \langle E_z \rangle$,  we find,
\begin{equation}
\frac{\langle E' \rangle}{\langle E \rangle} = \frac{1}{ (1+ \tilde{m})^2}   + \frac{ \tilde{m}^2}{9 (1+\tilde{m})^2 } \sum_{j,k \in (x,y,z)} \frac{   c_{0,j}^2 W_k^2 \beta_j^2}{ c_{0,k}^2 W_j^2 \beta_k^2   } \mathcal{M}_j[ (\mathrm{ce}_k(\tau)^2  + \mathrm{se}_k(\tau)^2 )^{-1}], 
\end{equation}
where $W_j = \mathrm{ce}_j(0) \dot{\mathrm{se}}_j(0)$ is the Wronskian and the operator $\mathcal{M}_j$ is defined by,
\begin{equation} \label{eq:mjOperator}
\mathcal{M}_j \left[ h(\tau)  \right] = \lim_{L\rightarrow \infty} \frac{1}{2L} \int^L_{-L}  h(\tau)   \left[\mathrm{ce_j}(\tau )^2+\mathrm{se_j}(\tau )^2\right]    d \tau.
\end{equation}
This ratio is plotted as a function of $\tilde{m}$ for two different values of $q_r = q_x = -q_y$ as a function of the mass ratio in Fig.~\ref{fig:energy_ratio_fullLocalisation}, finding excellent agreement with the results of numerical simulations of a collision (see Sec. \ref{section:numericalSimulations}) occurring under these conditions for two different values of $q_r$. 

At sufficiently high mass ratios, $\langle E' \rangle > \langle E \rangle$, that is, the ion has gained energy from the collision despite the fact that the collision has occurred at the exact centre of the trap with a buffer gas of zero kinetic energy. This occurs through a different mechanism than that investigated in Ref.~\cite{cetina12a} in that it does not require a non-zero interaction time and is independent of the magnitude of the ion-neutral interaction potential. It is due to the fact that the velocity associated with the micromotion does not vanish entirely for an ion with a non-zero amplitude of motion passing through the centre of the trap, see Appendix~\ref{section:centralCollisionAppendix}. Furthermore, as the energy gained in the collision is proportional to the pre-collision energy of the ion, this implies that repeated collisions lead to the  energy steadily increasing over time, that is, runaway heating and loss of the ion. 

The value of the mass ratio at which this occurs is significantly larger than the typical critical mass ratio of $\tilde{m} \approx 1.2$ for cooling with a uniform buffer gas, and for $q_r = 0.1$ the required value of $\tilde{m} \approx 592$ is large enough that runaway heating is prevented for all typically used pairs of ions and buffer gas particles. For $q_r = 0.5$, a mass ratio of $\tilde{m} = 16$ is sufficient to cause runaway heating even if collisions take place only at the centre of the trap. While this is a greatly extended range compared to a uniform buffer gas, it confirms that low Mathieu stability parameters should be used if possible. A more exact value for the central-collision critical mass ratio which relaxes the requirement that equipartition of energy holds, i.e., allows that $\langle E_j \rangle \neq \langle E_k \rangle$ for $j \neq k$, can be calculated using the method detailed in Appendix~\ref{section:centralCollisionAppendix}, which for $q_r = 0.1$ and $0.5$ produces $\tilde{m} = 593$ and $\tilde{m} = 17$ respectively, in good agreement with the values predicted above. 

\subsection{Buffer gas in a harmonic trap}
\label{section:harmonicbuffergas}
Next, we consider the more realistic case in which the buffer gas has a non-uniform density and a finite width. We assume that the density of the buffer gas depends only on the secular position of the ion, $r_j = A_j \cos (\tilde{\phi_j})$, where $\tilde{\phi_j} = \phi_j + \omega_j t$ is the instantaneous secular phase. This requires that the density of the buffer gas is essentially constant over the amplitude of the micromotion, that is, the density does not change significantly over the interval from $r_j(1 - q_j/2)$ to $r_j(1 + q_j/2)$. We further assume that the buffer gas is confined in a harmonic potential, such that the density of the buffer gas is described by a three-dimensional Gaussian distribution,
\begin{equation} \label{eq:bufferGasGaussian}
\rho(\mathbf{r})=\rho_x(r_x) \rho_y(r_y) \rho_z(r_z).
\end{equation}
Here,
\begin{equation}
\rho_j(r_j) =\frac{1}{\sqrt{2 \pi } \sigma_j} e^{-\frac {r_j^2}{2 \sigma_j^2} },
\end{equation}
with the width expressed in terms of the  temperature of the buffer gas $T_b$ and the harmonic trapping frequency $\omega_{j,b}$ as,
\begin{equation} \label{eq:neutralSigmaDef}
\sigma_j^2 = \frac{k_B T_b}{m_b \omega_{j,b}^2} . 
\end{equation}
For a typical buffer-gas mass of $80~$amu, temperature $1~\mu$K, and trapping frequency $\omega_{j,b} = 100 \times 2 \pi~$Hz, we find $\sigma_{j} \approx 16~\mu$m.

\begin{figure}[ht]
\centering
\includegraphics[width=.7\linewidth]{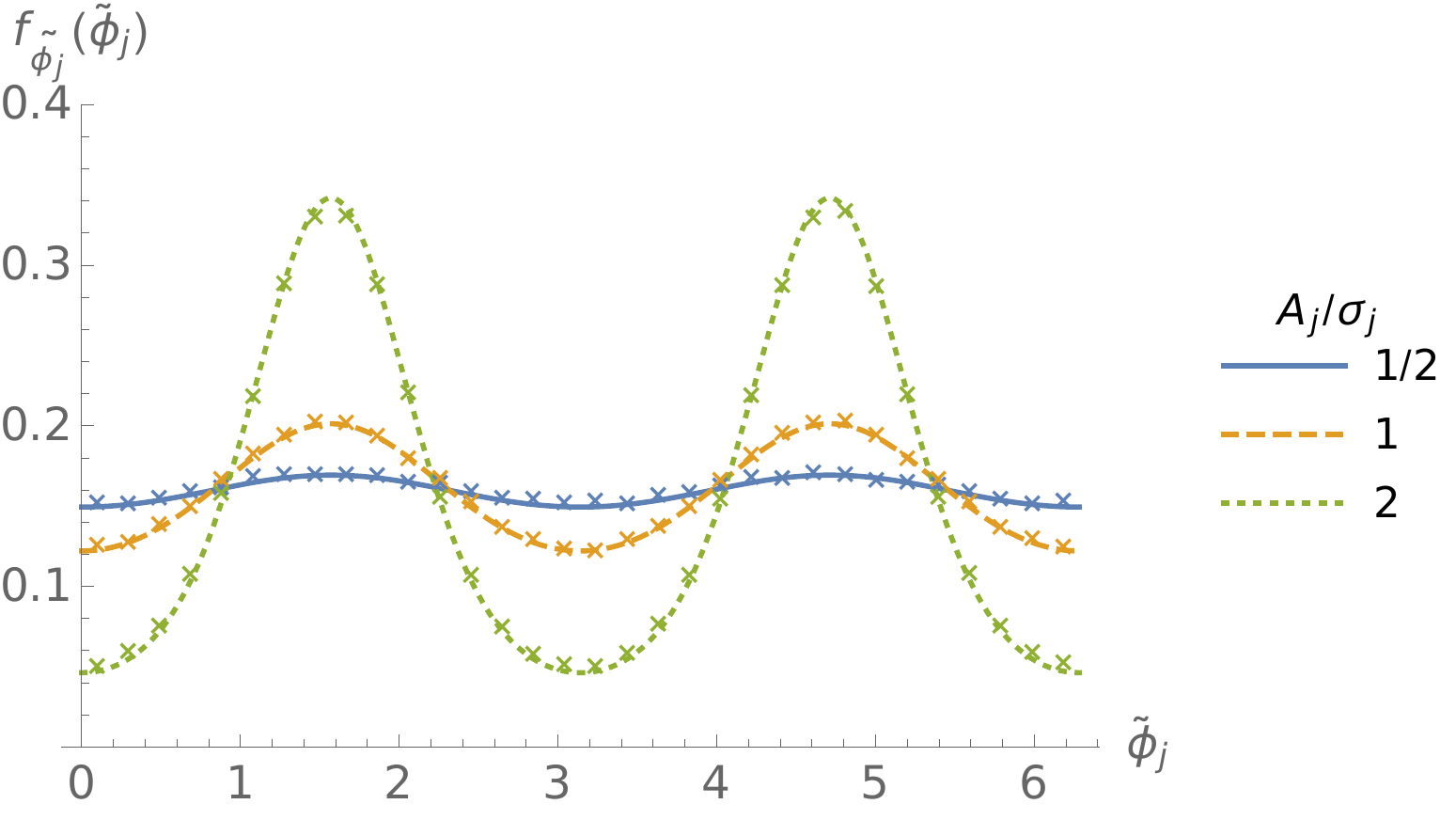}
\caption{Distribution of the instantaneous phase $\tilde{\phi}_j$ at the time of a collision as the result of buffer-gas localisation. The lines indicate the distribution of $\tilde{\phi}_j$ at the time of collision predicted using the analytical model described in the text, while the points indicate the distributions extracted from numerical simulations of collisions in a localised buffer gas. Three ratios of the amplitude of motion $A_j$ to the full width at half maximum of the density distribution of the buffer gas $\sigma$ are plotted. 500'000 simulations are run for each of the three cases. To ensure a well-defined value of the ratio $A_j/\sigma_j$, the numerical simulations are performed using a one-dimensional model of the motion by setting the initial amplitude for the motion along one axis to a constant relative to $\sigma_j = 1~$ (arb. units) and the amplitude of the remaining two axes equal to zero, with the Mathieu parameters for the axis with a non-zero amplitude of motion given by $q_j = 0.24, a_j = -0.00036$. The time at which a collision occurs is sampled using the method of Ref.~\cite{zipkes11a}, and the value of $\tilde{\phi}_j$ recorded and binned to produce the numerical distributions.}
\label{fig:secularPhaseAtCollision}
\end{figure}

Under these assumptions, the distribution for $\tilde{\phi_j}$ given that a collision has taken place, denoted $\tilde{\phi_j}|c$, is derived in Appendix~\ref{eq:appendix:secularPhaseDist}. The result is,
\begin{equation} \label{eq:phiDistLocalised}
f_{\tilde{\phi_j}}(\tilde{\phi_j} | c)=\frac{1}{2\pi}\frac{e^{-\frac{A_j^2 \cos (2 \tilde{\phi_j})}{4 \sigma_j^2}}}{  I_0\left(\frac{A_j^2}{4 \sigma_j^2}\right) } .
\end{equation}
where $I_n(x)$ is the modified Bessel function of the first kind \cite{olver2010a}. In Fig.~\ref{fig:secularPhaseAtCollision}, Eq.~\eqref{eq:phiDistLocalised} is plotted for three values of the ratio $A_j/\sigma_j$, finding good agreement with the results found from Monte-Carlo simulations of collisions employing the method of Ref.~\cite{zipkes11a} to bias the collision probability according to the density, then extracting the value of $\tilde{\phi_j}$ at the time of a collision. This distribution can be seen to vary from an effectively uniform distribution when $A_j = \frac{1}{2} \sigma_j$, to a sharply peaked distribution when $A_j = 2 \sigma_j$. That is, when the amplitude of the motion of the ion is small compared to the characteristic length scale of the buffer-gas density distribution, there is essentially no change. However, as the buffer gas density distribution becomes increasingly strongly localised around the centre of the trap, it is much more likely for $\phi_j$ to take values which correspond to the ion being at the centre of the trap at the time of a collision, $\tilde{\phi_j} =  \pi/2, 3\pi/2$.  

We may now evaluate the effects of localisation on the change in energy during a collision. As described above, this leads to a change in the distributions of the variables $\phi_j$, and since $\eta$ depends on these variables this leads to a change in the distribution of $\eta$ and thus its mean value. In particular, since $\phi_j$ depends on the energy in the presence of a non-uniform buffer gas, this implies that $\eta$ in turn becomes energy dependent. We obtain a representation of this effect by averaging $\eta$ over the distributions for the secular phases in the presence of a localised buffer gas, that is, by substituting $\phi_j \rightarrow \tilde{\phi}_j - \beta_j \tau$ into the definition of $\eta$ and integrating over $f_{\tilde{\phi}_j}(\tilde{\phi}_j)$. By symmetry, any integrals which are linear in $\cos {\phi}_j$ or $\sin {\phi}_j$ will vanish. However, the components of $\eta$ contain terms proportional to $\cos^2 {\phi}_j$ and $\sin^2 {\phi}_j$ which lead to non-zero integrals,
\begin{equation}
\int \cos^2(\tilde{\phi}_j- \beta \tau) f_{\tilde{\phi}_j} (\tilde{\phi}_j) d \tilde{\phi}_j = \frac{1}{2}-\frac{1}{2}\frac{I_1\left(\frac{A_j^2}{4 \sigma_j^2}\right)}{  I_0\left(\frac{A_j^2}{4 \sigma_j^2}\right) }\cos(2 \beta \tau).
\end{equation}
Since $A_j^2$ is proportional to $E_j$, the effect of a localised gas is to render $\eta$ dependent on the secular energy, as expected. The ratio of the Bessel functions may be expanded as $I_1(x)/I_0(x) \approx x/2$ for $x \lesssim 1/2$ \cite{olver2010a}, resulting in a term which is linear in $E_j$,
\begin{equation}\label{eq:besselRatioExpansion}
\frac{I_1\left(\frac{A_j^2}{4 \sigma_j^2}\right)}{  I_0\left(\frac{A_j^2}{4 \sigma_j^2}\right) } \approx \frac{E_j \tilde{m} \omega_{j,b}^2 }{  4  k_B T_b      \omega_j^2 c_{0,j}^2 } ,
\end{equation}
where Eq.~\eqref{eq:neutralSigmaDef} has been used to replace $\sigma_j$ by the buffer gas temperature and trapping frequency.  Consequently, when the secular energy is large, the value of $\eta$ is on average reduced compared to its value when $E$ is small or when the collisions take place in a uniform buffer gas. 

To incorporate this effect, we take a recurrence relation for $E'$ of the form,
\begin{equation} \label{eq:localisationLinearEta}
E' \approx \eta_0 E - \eta_1 E^2 + \epsilon = (\eta_0 - \eta_1 E) E + \epsilon,
\end{equation}
where $\eta_0$ is equivalent to $\eta$ in the absence of localisation, and $\eta_1 > 0$ represents the effects of localisation, which reduces the energy gained in each collision. For this approximation to remain valid, it is necessary that $A_j^2/(4\sigma_j^2) \lesssim 1/2$ such that Eq.~\eqref{eq:besselRatioExpansion} holds. Converting $A_j^2$ to the secular energy using Eq.~\eqref{eq:chapter6EnergyDef} and replacing $\sigma_j$ with the buffer gas temperature and frequency using Eq.~\eqref{eq:neutralSigmaDef}, this can be expressed as,
\begin{equation} \label{eq:energyConstraint}
E_j < \frac{1}{\tilde{m}} k_B T_b \frac{  \omega _j^2}{\omega_{j,b}^2  }.
\end{equation}
Taking typical neutral trap frequencies $\approx~100~$Hz, ion trap frequencies $\approx~1~$MHz, and mass ratios of order unity, this holds for $E_j/(k_B T_b) < 10^8$, i.e., for most realistic situations in which the present treatment is valid.  We assume that $E$ remains sufficiently low at all times that this criterion is met and so the expansion of $\eta \rightarrow \eta_0 - \eta_1 E$ remains valid.  

For this work, we are primarily interested in the mean value of $\eta_1$, i.e., $\langle \eta_1 \rangle$. This value requires averaging over all the collision parameters, and an explicit expression for this is given in Appendix~\ref{section:eta1Expression}. The dependence of $\langle \eta_1 \rangle$ on the various parameters is complicated, but the overall scaling behaviour is given by,
\begin{equation} \label{eq:averageEta1Approx}
\langle \eta_1 \rangle \propto \frac{1}{k_B T_b} \sum_{j \in (x,y,z)} \alpha_j \frac{\omega_{j,b}^2}{\omega_j^2} ,   
\end{equation}
where the set of values $\alpha_j,  j \in (x,y,z)$  depend on the mass ratio, on the Mathieu parameters (and hence indirectly on $\omega_j$) and on $\theta_\rho,\phi_\rho$. Thus, the impact of localisation is increased by decreasing the temperature of the buffer gas or by increasing the stiffness of the potential used to confine the buffer gas, both of which imply decreasing the extent of the buffer gas.   
\subsection{The energy distribution of an ion in a localised buffer gas}
We now address the issue of obtaining the steady-state energy distribution for an ion interacting with a localised buffer gas, focusing on the case of a buffer gas in a harmonic potential. Previously \cite{rouse17a}, we have demonstrated that the energy distribution in the case of a uniform buffer gas can be obtained using the formalism of superstatistics \cite{beck01a,beck03a}. In this formalism, the energy distribution is represented in terms of a superposition of thermal distributions, that is, an average of Eq.~\eqref{eq:energyDistConditional} over all possible values for the temperature, 
\begin{equation}
f_E(E) = \int f_E(E|T) f_T(T) d T, 
\end{equation}
where $f_T(T)$ is a distribution describing the probability for the fluctuating temperature to take a certain value. When $f_T(T)$ follows an inverse-Gamma distribution, then evaluating this integral produces Tsallis statistics exactly \cite{beck03a}. In Ref.~\cite{rouse17a}, we demonstrated that when $\eta$ is independent of $E$, the recurrence relation given by Eq.~\eqref{eq:energyRecurrence} can be mapped onto an equivalent recurrence relation for the temperature $T$,
\begin{equation}
T' = \eta T + \kappa T_b,
\end{equation}
where $\kappa T_b = \langle \epsilon \rangle/(3 k_B)$, with $\kappa$ given by \cite{rouse17a},
\begin{equation} \label{eq:kappaDef}
\kappa = \frac{\tilde{m}}{3 (1 + \tilde{m})^2} \sum_{j \in (x,y,z)} \frac{c_{0,j}^2 \beta_j^2}{W_j^2}.
\end{equation}
Solving this recurrence relation produces inverse-Gamma statistics for $T$ \cite{biro05a,rouse17a}, and thus Tsallis statistics for $E$. In the present case, $\eta$ is not independent of $E$, but we have demonstrated in the previous section that it can be expanded in terms of an energy-independent term $\eta_0$ and a term linearly proportional to the energy, $\eta_1 E$.    Based on the relation between the moments of the energy and the temperature derived in Appendix~\ref{section:superstatMoments}, this implies that the temperature recurrence relation contains a term proportional to $T^2$. Thus, we take a model of the form:
\begin{equation} \label{eq:localistionTempRecurrence}
T' = (\eta_0 - 4 k_B \langle \eta_1 \rangle  T) T + \kappa T_b,
\end{equation}
where the proportionality factor of $4 k_B$ arises from the relation between $\langle E^2\rangle$ and $\langle T^2 \rangle$ (Appendix~\ref{section:superstatMoments}). Since localisation only becomes significant at high energies (temperatures), the bulk of the distribution is only weakly sensitive to this effect, and so the fluctuations in $\eta_{1}$ are neglected to ensure that there is only one source of noise present.  The procedure for solving this recurrence relation is detailed in Appendix~\ref{section:besselTsallisDerivation}, and the resulting energy distribution is given by,
\begin{equation} \label{eq:besselTsallisDist}
f_E^{(BT)}(E) =  \left(\frac{b}{\nu  E_\ell}\right)^{\frac{3}{2}} \frac{E^2}{\left(\frac{b E}{\nu }+1\right)^{\frac{3+\nu }{2} }} \frac{   K_{3+\nu}\left(\sqrt{   \frac{ E}{E_\ell }+\frac{\nu}{b E_\ell}   }\right)}{16 K_{\nu }\left(\sqrt{\frac{\nu }{b E_\ell}}\right)   },
\end{equation}
where $K_y(z)$ is the modified Bessel function of the second kind \cite{olver2010a} and we have assumed $k=2$ which is appropriate for a three-dimensional harmonic oscillator \cite{pethick01a}. The parameters for this distribution are defined in terms of $\mu = \langle \ln \eta_0 \rangle$ and $\sigma^2~=~\langle (\ln \eta_0)^2 \rangle - \mu^2$ as,
\begin{equation}
b = \frac{-\mu}{\kappa k_B T_b} ,
\end{equation}
\begin{equation}\label{eq:nuDef}
\nu = \frac{- 2 \mu}{\sigma^2},
\end{equation}
\begin{equation}
E_\ell = \frac{\sigma^2}{32 \langle \eta_1 \rangle}.
\end{equation}
In practice, direct calculation of these parameters is prevented due to the lack of analytical expressions for the required averages of $\ln \eta_0$. Thus, the parameters $b,\nu,E_{\ell}$ are most conveniently found numerically, as presented in Section~\ref{section:results} and discussed further in Appendix~\ref{section:parameterEstimation}.

Eq.~\eqref{eq:besselTsallisDist} is normalisable as long as $E_\ell >0$ and the ratio $b/\nu$ is positive, i.e., both $b$ and $\nu$ have the same sign, and has the form of a gamma-generalised inverse Gaussian distribution \cite{gomez13a}. Here, we refer to it as a Bessel-Tsallis (BT) distribution, reflecting the fact that the asymptotic behaviour for large energies is determined by the Bessel function and that it reduces to a Tsallis distribution in the limit $E_\ell \rightarrow \infty$, as we now demonstrate.  When the argument of the Bessel function is close to zero, i.e., if $E_\ell >> E, \nu/b$, the function may be approximated by \cite{olver2010a},
\begin{equation}
K_y(z) \approx \frac{1}{2}\Gamma(y) \left(\frac{z}{2}\right)^{-y}  ,
\end{equation}
where $\Gamma(y)$ is the Gamma function. By replacing the Bessel functions in Eq.~\eqref{eq:besselTsallisDist} with this approximation and simplifying the result, we find,
\begin{equation}
f_E^{(BT)}(E) \approx   \frac{E^2 b^3 \Gamma(\nu+3) }{2 \Gamma(\nu) }   \left(\frac{1}{b E/\nu+1}\right)^{\nu+3},
\end{equation}
which is equivalent to Eq.~\eqref{eq:tsallisEnergyDistFirst} with $k = 2$, $b = \langle \beta \rangle$ and $\nu = n_T$ \footnote{In fact, $b$ and $\nu$ are only estimates for $\langle \beta \rangle, n_T$ due to the approximations made in the derivation of this distribution, see Refs.~\cite{sornette97a,rouse17a}.}. Thus, when $E_\ell \rightarrow \infty$, the distribution converges to Tsallis statistics. This corresponds to $\langle \eta_1 \rangle \rightarrow 0$, i.e., the effects of localisation becoming negligible for a uniform buffer gas. More generally, for values of  $E << E_\ell$, the distribution closely resembles Tsallis statistics.  

We also consider the opposite limit, i.e., large values of the energy relative to $E_\ell$. For large values of the argument the Bessel function asymptotically approaches the form,  \cite{olver2010a},
\begin{equation}
K_{3+\nu}\left(\sqrt{\frac{E}{E_\ell}+\frac{\nu }{b E_\ell}}\right) \sim e^{-\sqrt{\frac{E}{E_\ell}+\frac{\nu }{b E_\ell}} } .
\end{equation}
For large values of $E$, $\nu/(b  E_\ell)$ is negligible, and thus Eq.~\eqref{eq:besselTsallisDist}   asymptotically approaches zero as  $e^{-\sqrt{E/E_\ell}}$.  This leads to a slower asymptotic decay than that of  a thermal distribution or the exponential-Tsallis distribution, but a faster decay than the power-law tail of the Tsallis distribution. The parameter $E_\ell$ controls the rate of this decay, with a small value of $E_\ell$, i.e., a large value of $\langle \eta_1 \rangle$, causing the distribution to more rapidly approach zero. $E_\ell$ therefore represents a characteristic energy above which the distribution deviates from Tsallis statistics. The exact dependence of $E_\ell$ on the trapping parameters is complex due to depending on both $\sigma^2$ and $\langle \eta_1 \rangle$, but in general $E_\ell \propto k_B T_b$, and decreases as the ratio $\omega_{j,b} / \omega_{j}$  increases. This reflects the fact that as the density of the buffer gas becomes increasingly strongly peaked at the centre, i.e., at lower temperatures or larger values of $\omega_{j,b}$, the effects of localisation become significant at lower energies. 
  
\subsection{Numerical simulations}\label{section:numericalSimulations}
To test the validity of the distribution derived in the previous section, we perform numerical simulations using the method of Ref.~\cite{devoe09a}, with the rate of collisions biased according to the density of the buffer gas as described in Ref.~\cite{zipkes11a}. In brief, given the position and velocity of the ion at a given time, the simulation is advanced a random amount of time chosen from an exponential distribution. The position and velocity at this time are calculated using the exact solutions to the Mathieu equations. The probability for a collision to occur is then determined based on the density of the buffer gas at the location of the ion (including the displacement due to the micromotion), and if a collision occurs the velocity is updated according to Eq.~\eqref{eq:collisionAtomVel1}. The simulation proceeds until a predetermined number of collisions have taken place.  When the energy of the ion is high, the ion typically occupies regions of space with a low probability for collisions to occur, and so the simulation must be advanced many times before a collision takes place.  Consequently, for the simulations performed here we are limited to mass ratios $\tilde{m} < 10$ and small values of the Mathieu stability parameters to ensure that the simulations do not take an intractable amount of computational time.  The velocity of the buffer gas is assumed to follow classical Maxwell-Boltzmann statistics, and the heating and depletion of the buffer gas due to the collisions with the ion is neglected. That is, the density and velocity distributions are assumed to remain fixed throughout the simulation, see Ref.~\cite{zipkes11a} for how the heating of the buffer gas may be implemented. The energy gain due to the finite ion-neutral interaction time is neglected \cite{cetina12a}.  Each iteration of a simulation consists of a series of 500 collisions, after which the final secular energy of the ion is recorded. For the simulations performed in this work, we assume an ideal harmonic linear rf trap with $q_x = 0.1, q_y = - q_x, q_z = 0$, $a_z = 0.00625$ and an rf frequency $\Omega = 20 \times 2 \pi~$MHz. The buffer gas is assumed to be confined in a radially symmetric potential with the axial frequency set equal to one-half of the radial buffer gas trapping frequency $\omega_{r,b}$, that is, $\omega_{x,b} = \omega_{y,b} = \omega_{r,b} , \omega_{z,b} = \frac{1}{2} \omega_{r,b}$. We assume for all simulations that no excess micromotion is present, that the centres of the two traps coincide, and that both traps share the same coordinate system.

\section{Results}\label{section:results}
In Section \ref{section:harmonicbuffergas}, we derived an analytical form for the energy distribution obtained for an ion immersed in a non-uniform buffer gas with the parameters expressed in terms of the change in energy during a single collision. We now must investigate both the form of the distribution and the accuracy of these parameters.

\begin{figure}[tb]
\centering
\includegraphics[width=.7\linewidth]{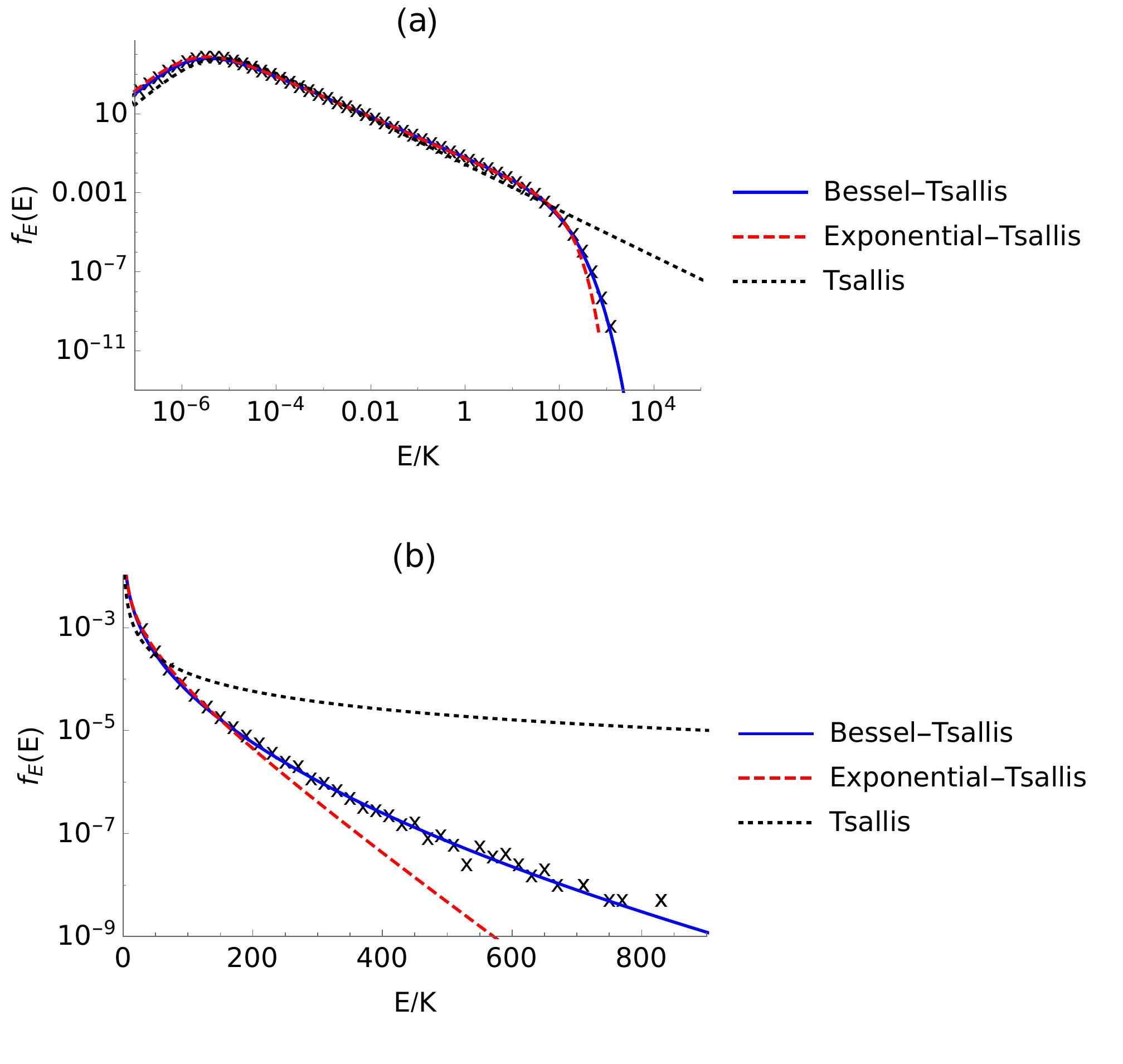}
\caption{(a) The secular energy distribution obtained for an ion in a linear rf trap interacting with a buffer gas confined in a harmonic potential, with a mass ratio $\tilde{m} = 2$, Mathieu parameters $q_r = 0.1, a_z = 0.000625$, rf frequency $\Omega = 20 \times 2 \pi~$MHz, buffer gas temperature $T_b = 1~\mu$K, and buffer-gas radial trap frequency $\omega_{r,b} = 100 \times 2 \pi$ Hz, with the axial frequency equal to half of this value. Points indicate the distribution obtained from 10'000'000 iterations of the numerical simulations. The lines indicate different analytical forms for the distribution discussed in the main text with their parameters found from fitting to the numerical data via maximum-likelihood estimation (Tsallis, Exponential-Tsallis) or from the collision model (Bessel-Tsallis).  As (a), but shown on a linear-logarithmic scale to highlight the difference in the tails of the distribution.}
\label{fig:LocalisationPaperDistributionsLinLog}
\end{figure}

 In Fig.~\ref{fig:LocalisationPaperDistributionsLinLog}, the Tsallis, exponential-Tsallis, and Bessel-Tsallis distributions are compared to the numerical data for $\omega_{r,b}=100~$Hz, $\tilde{m}=2$. The parameters for the Bessel-Tsallis distribution can be calculated numerically from the collision model as discussed in Appendix~\ref{section:parameterEstimation}, while the parameters for the standard Tsallis and exponential-Tsallis distributions are found by fitting the distribution to the numerical data via maximum-likelihood estimation. It can be seen that the low-energy behaviour follows Tsallis statistics as expected, with a noticeable disagreement at higher energies. The Bessel-Tsallis distribution appears a better fit than the exponential-Tsallis distribution. This is more apparent when plotting the distributions on a scale in which $E$ is shown linearly and $f_E(E)$ is plotted logarithmically, see Fig.~\ref{fig:LocalisationPaperDistributionsLinLog}(b). The simulations performed here more realistically include an axial component for the trapping potential of the ion and do not make the adiabatic approximation, and so are not equivalent to those performed in Ref.~\cite{hoeltkemeier16a} for which the exponential-Tsallis distribution was found to be a good fit to the data.

\begin{figure}[tb]
\centering
\includegraphics[width=.7\linewidth]{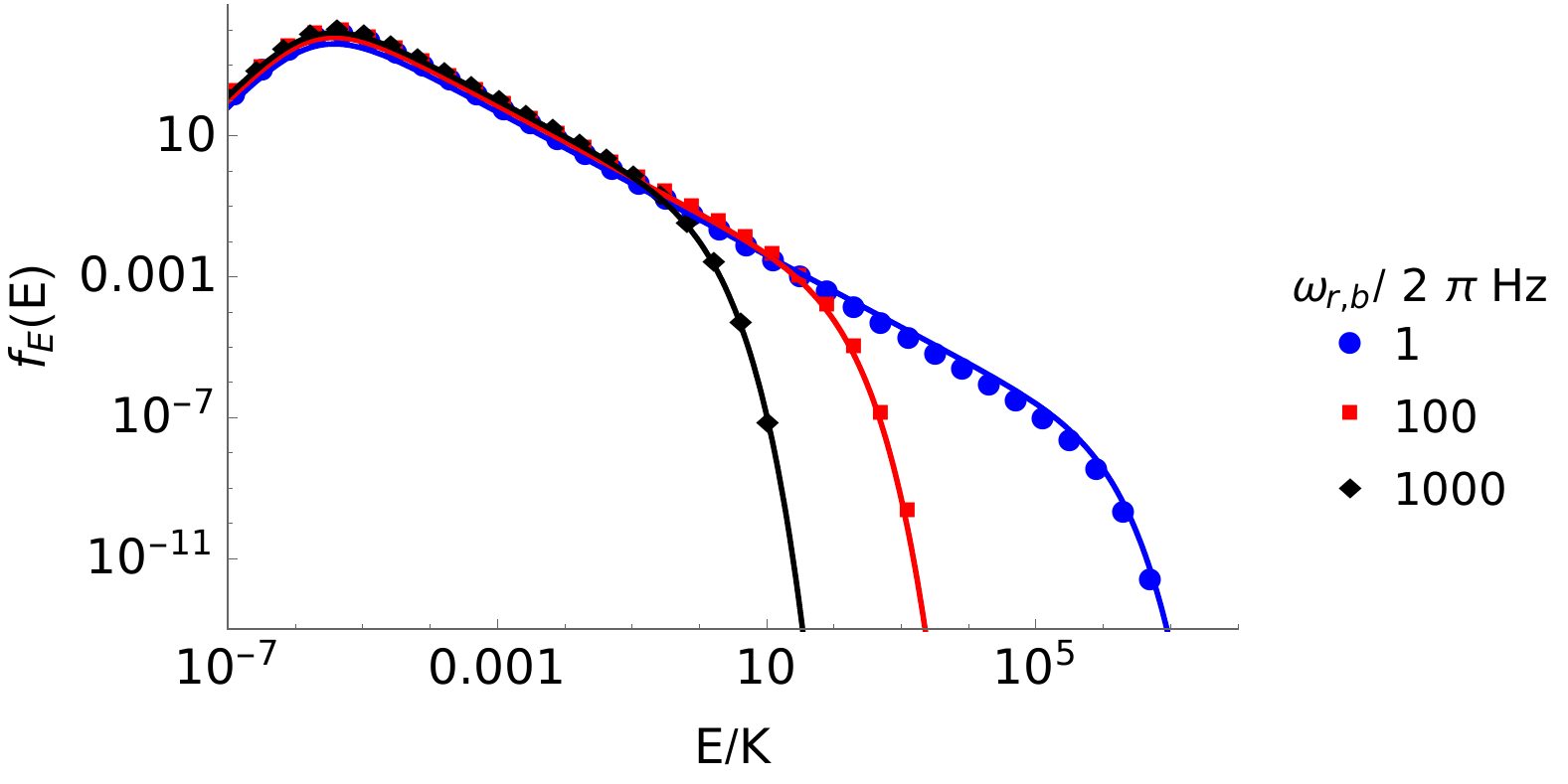}
\caption{The energy distribution obtained for an ion in a linear rf trap interacting with a buffer gas confined in a harmonic potential. The mass ratio $\tilde{m} = 2$, Mathieu parameters $q_r = 0.1, a_z = 0.000625$, rf frequency $\Omega = 20 \times 2 \pi~$MHz and buffer gas temperature $T_b = 1~\mu$K were kept fixed, whereas the radial trapping frequency for the buffer gas was varied over three different values, and the axial frequency for the neutral buffer gas was set to one half of this value. 10'000'000 simulations of 500 collisions were performed for each of the three cases shown, and the resulting energy distributions were log-binned resulting in the distributions shown by the points. The lines show the Bessel-Tsallis distribution derived in the main text, with the parameters calculated by numerically from the change in energy during a collision. }
\label{fig:LocalisationPaperDistributions}
\end{figure}

We repeat the simulations at this mass ratio with the radial frequency of the neutral trap set to values of $\omega_{r,b} = (1, 1000) \times 2 \pi~$Hz to investigate the change of the distribution over the range of trapping frequencies typically achievable using magnetic or magneto-optical traps for neutral particles. These distributions and that for $\omega_{r,b} = 100 \times 2 \pi$~Hz are compared in Fig.~\ref{fig:LocalisationPaperDistributions}.  The Bessel-Tsallis distribution (Eq.~\eqref{eq:besselTsallisDist}) with parameters found from the collision model is found to accurately describe these distributions over multiple orders of magnitude, confirming the validity of the analytical model. We note also that the values of $E_\ell$ found numerically obey the scaling relation with respect to $\omega_{r,b}$ predicted from Eq.~\eqref{eq:averageEta1Approx}, that is, $E_\ell \propto 1/\omega_{r,b}^2$. Thus, increasing the trapping frequency from $\omega_{r,b} = 100 \times 2 \pi~$Hz by a factor of 10 was found to reduce $E_\ell$ by a factor of 100.

\begin{figure}[tb]
\centering
\includegraphics[width=.6\linewidth]{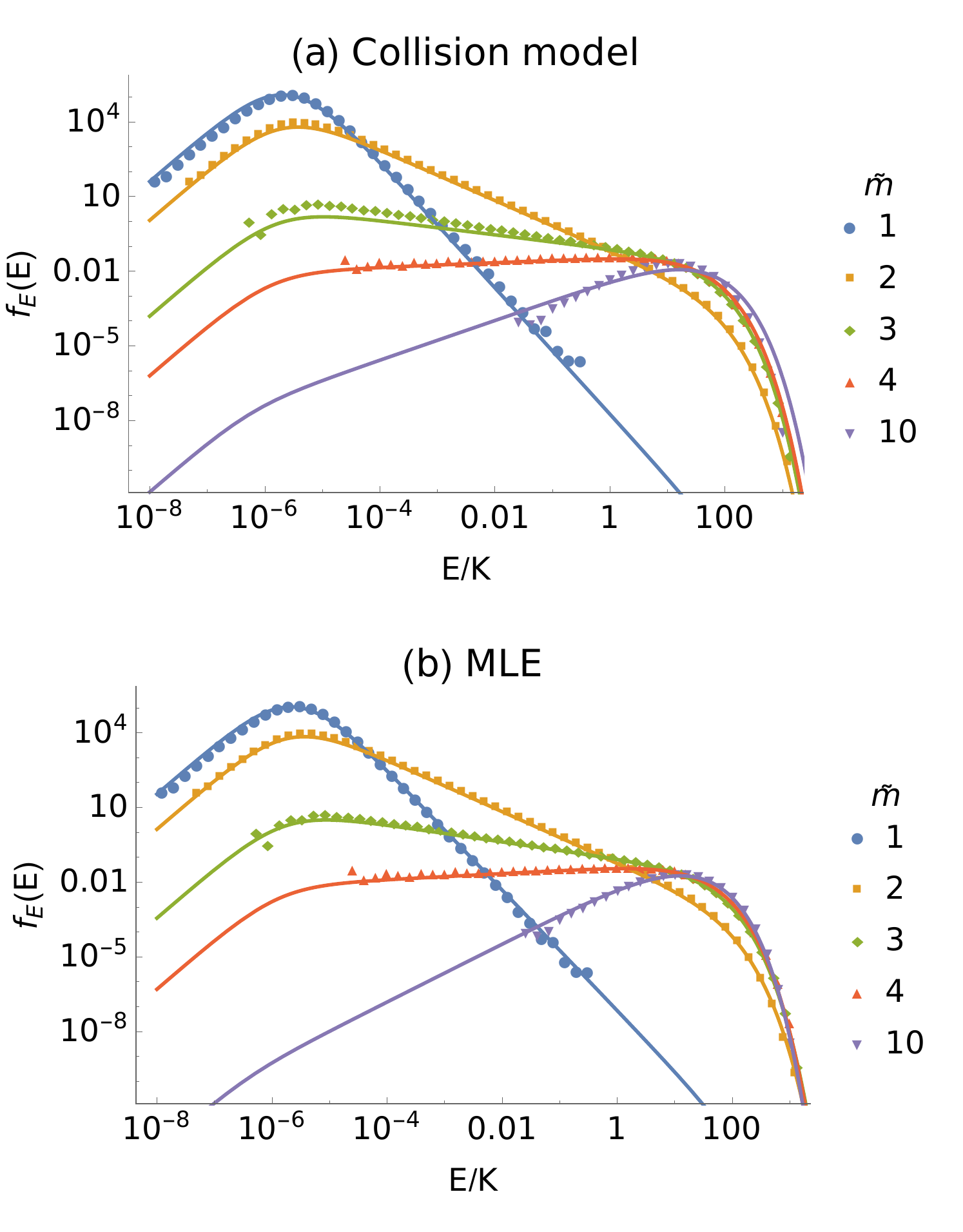}
\caption{(a) The energy distribution obtained for an ion in a linear rf trap interacting with a buffer gas confined in a harmonic potential. The Mathieu parameters $q_r = 0.1, a_z = 0.000625$, rf frequency $\Omega = 20 \times 2 \pi~$MHz, buffer gas temperature $T_b = 1~\mu$K and trapping potential for the buffer gas were kept fixed, while the buffer gas-to-ion mass ratio $\tilde{m}$ was set to different values in the interval $1-10$.  10'000'000 simulations of 500 collisions were performed for each of the distributions with $\tilde{m} = 1-4$, with $\approx 4'000'000$ simulations performed for $\tilde{m}=10$ due to the extended computational time required at high mass ratios. The resulting energy distributions are log-binned resulting in the distributions shown by the points.  The lines give the Bessel-Tsallis distribution derived in the main text, with the parameters extracted from the change in energy during a collision. (b) As (a), but the parameters for the Bessel-Tsallis distribution are found by fitting the distribution to the numerical data via maximum-likelihood estimation (MLE).}
\label{fig:LocalisationDistMTilde}
\end{figure}

It is also of interest to vary the mass ratio while keeping the trapping frequencies fixed ($\omega_{r,b} = 100\times 2 \pi$ Hz and the same Mathieu parameters as before), with the results obtained for $\tilde{m} = 1-10$ shown in Fig.~\ref{fig:LocalisationDistMTilde}. For $\tilde{m} = 1$, the distribution is concentrated around the thermal energy of the buffer gas, $E \approx 1~\mu$K, whereas for large mass ratios $\tilde{m} > 4$, the peak of the distribution appears at much higher energies. This behaviour may be explained using a random-walk analogy as in Ref.~\cite{sornette97a}. When $\tilde{m}$ is small,  $\langle \eta \rangle << 1$, and the energy drifts towards smaller values before being reflected off a barrier at the thermal energy of the buffer gas, $\approx k_b T_B$. Conversely, for large values of the mass ratio, $\langle \eta \rangle > 1$, and the energy drifts towards large values but is prevented from reaching larger values due to the effects of localisation,  which is analogous to the existence of a barrier at $\approx E_\ell$. In both of these cases, the distribution is peaked around the location of this barrier. For intermediate mass ratios, $\langle \eta \rangle \approx 1$ and there is no strong drift in either direction, leading to the energy distribution spreading out between these two barriers. This produces the distribution seen spanning multiple orders of magnitude for $\tilde{m} = 2$ and $3$.  In all cases, the numerical distributions are adequately reproduced by the Bessel-Tsalis distribution with parameters found from numerical sampling of $\eta$, confirming the validity of the model. There is a slight disagreement in terms of the slope of the distribution in the region between $E \approx k_B T_b$ and $E \approx E_\ell$, particularly for $\tilde{m}=3$ and $10$. In this regime, the distribution is primarily characterised by $\nu$. In the absence of localisation, i.e., for a model of the form $E' = \eta E + \epsilon$ where $\eta$ is independent of $E$, the value of $\nu$ depends on the distribution of $\eta$ and $\epsilon$ \cite{rouse17a,rouse18a,sornette97a,buraczewski16a}, and is only approximated by Eq.~\eqref{eq:nuDef}. It is reasonable to assume that this also applies in the localised case, such that the true values of the parameters deviate from the estimates given here. Indeed, the agreement is improved when the set of values $b,\nu,E_\ell$ are found through fitting the Bessel-Tsallis distribution to the numerical data through maximum-likelihood estimation, as can be seen in Fig.~\ref{fig:LocalisationPaperDistributions}(b). From this, we conclude that the functional form of the distribution is correct over a wide range of mass ratios, and the definitions of $b,\nu,E_\ell$ given by Eqs.~(34-36) serve as good approximations to the actual values of these parameters.

Finally, let us briefly comment on the limits of the validity of the distribution derived here. The Bessel-Tsallis distribution is obtained under the assumption that the trajectory of the density of the buffer gas is constant over the amplitude of the micromotion of the ion requiring small values of $q_j$. Moreover, it is assumed that the energy of the ion remains sufficiently low such that Eq.~\eqref{eq:energyConstraint} holds. For realistic experimental conditions, i.e., low mass ratios, a linear rf trap with $q_x = -q_y = 0.1$ and harmonic buffer-gas trapping frequencies on the order of $100~$Hz, these conditions are met. In Section~\ref{section:centralCollisions}, however, it was shown that for a sufficiently large mass ratio the energy of the ion may increase on average with each collision even if all collisions take place at the trap centre. For a buffer gas with a finite width, it is reasonable to assume that this begins at a mass ratio between the value obtained in Section~\ref{section:centralCollisions} for a buffer gas limited to the centre of the trap and the value of $\tilde{m}=1.2$ found for a uniform buffer gas. In this case, the energy of the ion will increase beyond the constraint given by Eq.~\eqref{eq:energyConstraint} and so deviations from the Bessel-Tsallis distribution are possible. Likewise, if the Mathieu $q_j$ parameters are much larger than the value of $q_j = 0.1$ typically used, then the assumption that the density of the buffer gas depends only on the secular position of the ion breaks down, and again it is possible that a different distribution is obtained. 

\section{Conclusions}
In this work, we have extended our analytical model of Ref.~\cite{rouse17a,rouse18a} for ion-neutral collisions in a rf trap to include the effects of a localised buffer gas. We have shown that this greatly enhances the range of mass ratios over which sympathetic cooling is possible in line with recent experimental and numerical findings \cite{dutta17a,hoeltkemeier16a,hoeltkemeier16b}, and found an upper limit to the mass ratio which may successfully cool an ion if all collisions are limited to the exact centre of the trap. Furthermore, we have derived an analytical form for the energy distribution observed when the buffer gas is held in a harmonic potential with trapping frequencies much smaller the secular frequencies of the ion trap. This energy distribution is found to converge to Tsallis statistics at low energies, but exhibits a high-energy tail with a decay of the form $E^{-\sqrt{E/E_{\ell}}}$, where $E_\ell$ is a characteristic energy scale. The analytical model is confirmed using numerical simulations, finding good agreement over a range of mass ratios and trapping frequencies for the neutral buffer gas when the parameters are estimated from the change in energy during a collision found from numerical simulations. The agreement is further improved when the distribution is fit to the energy distributions via maximum-likelihood estimation. 

\section*{Acknowledgments}
We acknowledge funding from the University of Basel, the Swiss Nanoscience Institute project P1214 and the Swiss National Science Foundation as part of the National Centre of Competence in Research, Quantum Science \& Technology (NCCR-QSIT) and grant nr. 200020\_175533. We thank Z. Meir for helpful discussions.
 
\bibliography{thesis_extra_refs,Main-May18}
\appendix

\section{Energy change during collisions at the centre of the trap} \label{section:centralCollisionAppendix}
\begin{figure}[tb]
\centering
\includegraphics[width=.5\linewidth]{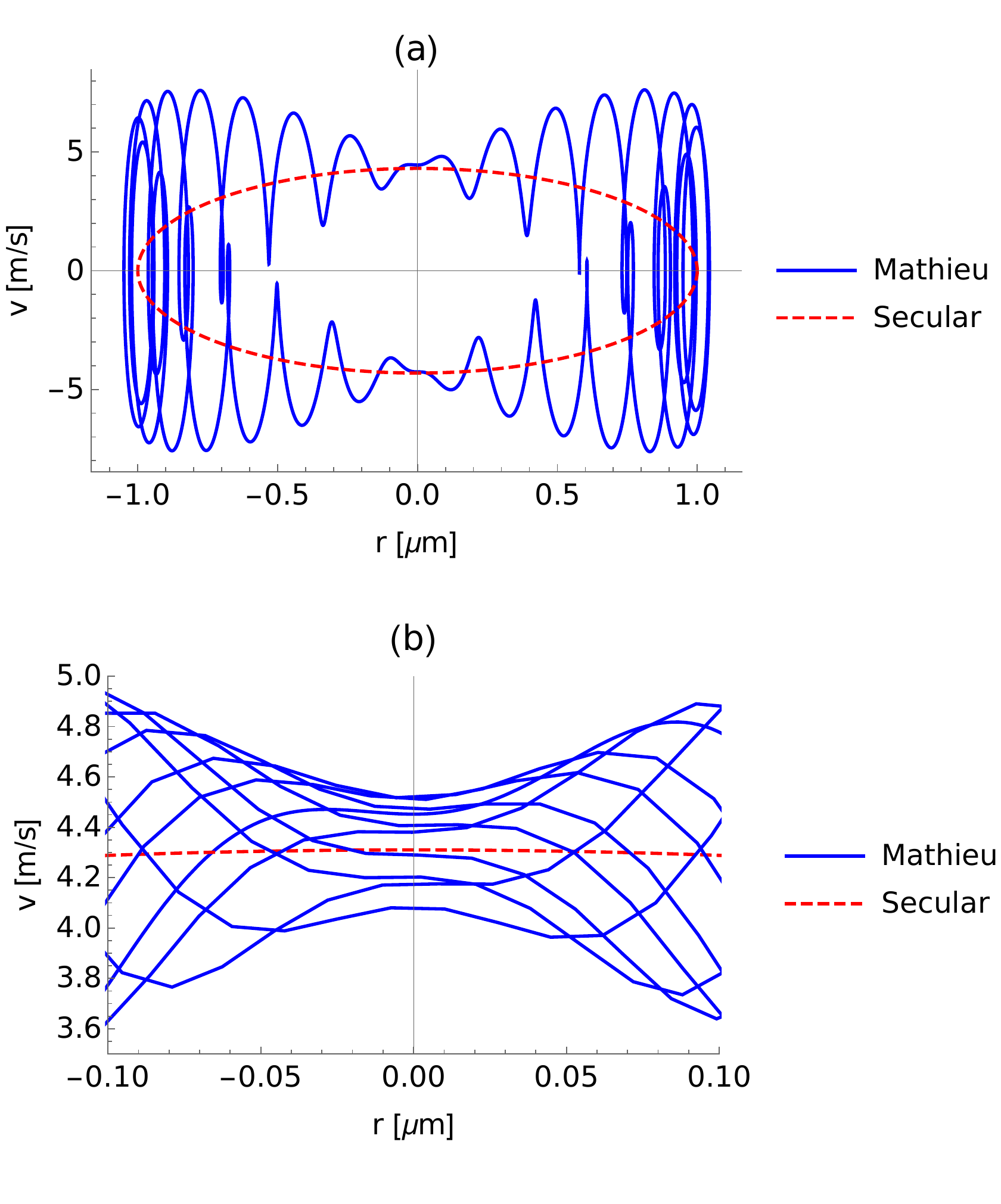}
\caption{The phase-space trajectory of an ion in a radiofrequency trap with $q_j = 0.1, a_j = -0.000625/2, \Omega = 20 \times 2 \pi$ MHz, an amplitude of $A_j = 1~\mu$m, and the phase set to $\phi_j = 0$. (a) The trajectory over one period of the secular motion comparing the exact solution to the Mathieu equation (solid line) to the secular motion (dashed line). (b) The velocity of the ion close to the centre of the trap, $r = 0$, shown for multiple periods of the secular motion to highlight the presence of micromotion at the centre of the trap.  }
\label{fig:LocalisationPaperMathieuTrajectory}
\end{figure}

We consider the extreme case in which collisions may only occur at the centre of the trap, under the assumption that there is no excess micromotion due to external forces (see, e.g., Ref.~\cite{berkeland98a}). This may be achieved using the same procedure to calculate the change in energy as described in Section~\ref{section:energyChangeInCollision}, with the exception that there is now an additional constraint that the collision must occur at the centre of the trap, $r_j(\tau) = 0$ for each $j \in (x,y,z)$, where $r_j(\tau)$ is defined as in Eq.~\eqref{eq:mathieuPositionHomogenous}. That is,
\begin{equation} \label{eq:phaseCollisionAtCentre}
 A_j (\cos \phi_j ~\matcj - \sin \phi_j ~\matsj) = 0.
\end{equation}
For non-zero values of $A_j$, corresponding to an ion with a non-zero amplitude of motion passing through the trap centre, solutions to Eq.~\eqref{eq:phaseCollisionAtCentre} can be found by requiring that,
\begin{equation} \label{eq:phijAtCentre}
\tan \phi_j = \frac{\matcj}{\matsj} .
\end{equation}
Since $\tan \phi_j$ is periodic, there are two possible solutions to this equation, which physically represent the fact that the velocity may correspond to motion in either the $+j$ or $-j$ direction. The velocity is given by the derivative of Eq.~\eqref{eq:mathieuPositionHomogenous} with respect to $\tau$, and substituting in the solutions given by Eq.~\eqref{eq:phijAtCentre} we obtain,
\begin{equation} \label{eq:velAtCentre}
v_j(\tau , r_j = 0) = \pm \frac{A_j W_j}{\sqrt{\matcj^2 + \matsj^2}} .
\end{equation}
Here, $W_j = \matcj \matsjp - \matcjp \matsj$ is the Wronskian, which is a time-independent quantity \cite{boyce17a}. For $q_j = 0$, $\matcj, \matsj$ reduce  to $\cos \beta_j \tau$ and   $\sin \beta_j \tau$ respectively, such that $\matcj^2 + \matsj^2 = 1$. In this case, the velocity at the centre of the trap is independent of $\tau$, as expected for a harmonic oscillator. However, for non-zero $q_j$ this no longer holds. Indeed, by plotting the phase-space trajectory of the ion (Fig.~\ref{fig:LocalisationPaperMathieuTrajectory}), it can be seen that the velocity of the ion at $r_j = 0$ is not equal to the secular velocity and takes a range of values. 

We estimate the magnitude of this effect as follows. Using the Fourier series definitions for $\matcj, \matsj$ and trigonometric identities, it can be shown that,
\begin{equation} \label{eq:mathieuSqSumApprox}
\mathrm{ce}_j(\tau)^2 + \mathrm{se}_j(\tau)^2  = \sum_{m,n} c_{2m,j} c_{2n,j} \cos[ 2(m-n) \tau].
\end{equation}
As a result of the terms in this sum with $m \neq n$, this function is time dependent and contains components with frequencies of integer multiples of $\Omega$.   Evaluating Eq.~\eqref{eq:mathieuSqSumApprox} for $m,n \in (-1,0,1)$, and using the approximate values for the Mathieu coefficients $c_0 = 1, c_{\pm 2,j} = -q_j/4$ \cite{olver2010a}, we find,
\begin{equation}
    \mathrm{ce}_j(\tau)^2 + \mathrm{se}_j(\tau)^2 \approx 1 - q \cos (2 \tau).
\end{equation}
The velocity is therefore approximated by,
\begin{equation} \label{eq:velAtCentre2}
v_j(\tau, r_j = 0) \approx \pm \frac{A_j W_j}{\sqrt{1 - q_j \cos (2 \tau)}} ,
\end{equation}
and expanding this as a Taylor series to first order in $q_j$ around $q_j = 0$ we obtain,
\begin{equation} \label{eq:velAtCentre3}
v_j(\tau, r_j = 0) \approx \pm  A_j W_j[ 1 + \frac{q_j}{2} \cos (2 \tau) ].
\end{equation}
An approximate value for the Wronskian $W_j$ can be found using the $m=0$ terms of the Fourier series expansions of $\matcj,\matsj$, which produces $W_j \approx c_0^2 \beta_j$. Using this approximation with $c_0 \approx 1$, and converting from $\tau$ to $t$, we find,
\begin{equation}
    v_j(t, r_j = 0) \approx \pm A_j \omega_j [ 1 + \frac{q_j}{2} \cos (\Omega t)], 
\end{equation}
where the definition of the secular frequency $\omega_j = \beta_j \Omega /2 $ has been employed to simplify the result. This form of the result bears a resemblance to the adiabatic approximation for the motion of the ion \cite{berkeland98a},
\begin{equation} \label{eq:motionAdiabatic}
    r_j(t) = A_j \cos(\omega_j t + \phi_j)[1 - \frac{q_j}{2} \cos (\Omega t)],
\end{equation}
where the sign of the $\frac{q_j}{2}$ term used here differs from that of Ref.~\cite{berkeland98a} as a result of the use of a different convention for the Mathieu equation. Taking the derivative of Eq.~\eqref{eq:motionAdiabatic} with respect to $t$ produces,
\begin{equation}
    v_j(t) = A_j \cos(\omega_j t + \phi_j)[ \Omega \frac{q_j}{2} \sin (\Omega t)] - A_j \omega_j \sin(\omega_j t + \phi_j)[1 - \frac{q_j}{2} \cos (\Omega t)].
\end{equation}
For the ion to be at the centre of the trap with $|q_j| < 2$ it is required that $\cos(\omega_j t + \phi_j) = 0$, and therefore $\sin(\omega_j t + \phi_j) = \pm 1$. Thus,
\begin{equation}
    v_j(t,r_j=0) = \mp A_j \omega_j [1 - \frac{q_j}{2} \cos (\Omega t)],
\end{equation}
which is equivalent to Eq.~\eqref{eq:velAtCentre3} up to the sign of the $\frac{q_j}{2}$ term. This discrepancy in the sign is a consequence of the fact that $\omega_j q_j$ is approximately proportional to $q_j^2$ \cite{olver2010a}, and terms of this order are not included in the adiabatic approximation. Using an improved approximation for the motion of the ion including terms up to order $q_j^2$ as in Ref.~\cite{wineland98a}  produces a result in agreement with Eq.~\eqref{eq:velAtCentre3}.

The velocity of the ion at the centre of the trap is therefore approximately given by the sum of the secular velocity and a term proportional to $q_j/2 \cos (\Omega t)$, i.e., a micromotion term. For the trajectory shown in Fig.~\ref{fig:LocalisationPaperMathieuTrajectory} with $q_j = 0.1$, the secular velocity is given by $\approx 4.3$ m/s, and the actual velocity spans $4.1 - 4.5$ m/s, in agreement with this result. Thus, we conclude that an ion which is passing through the centre of the trap exhibits a contribution to the velocity from the micromotion proportional to the secular velocity of the ion. If the ion is perfectly cooled to the centre of the trap, then $A_j =0$ and this micromotion vanishes, as expected. However, for non-zero values of $A_j$, the magnitude of this micromotion increases proportionally to the secular velocity. 

The post-collision energy is given as before by Eq.~\eqref{eq:energyComponents}, as this expression is valid for collisions at an arbitrary point in the trap. However, in this case the set of phases $\phi_j$ are determined by Eq.~\eqref{eq:phijAtCentre}, such that each $\dist{\phi_j}$ is sharply peaked at the two possible values which we assume to occur with equal probability. Assuming a thermal distribution for the components of the velocity of the buffer gas, an isotropic random rotation matrix, and taking $\tau$ to follow a uniform distribution, we find,
\begin{equation} \label{eq:centralCollisionEnergy}
 \langle E'_j  \rangle = \frac{\langle E_j \rangle}{ (1+ \tilde{m})^2} + \kappa_j k_B T_b + \sum_{k\in (x,y,z)} \frac{ \tilde{m}^2 c_{0,j}^2 W_k^2 \beta_j^2}{3 (1+\tilde{m})^2 c_{0,k}^2 W_j^2 \beta_k^2   } \mathcal{M}_j[ (\mathrm{ce}_k(\tau)^2  + \mathrm{se}_k(\tau)^2 )^{-1}] \langle E_k \rangle,
\end{equation}
where, 
\begin{equation} \label{eq:kappaJ}
\kappa_j = \frac{\tilde{m}}{  (1 + \tilde{m})^2}  \frac{c_{0,j}^2 \beta_j^2}{W_j^2},
\end{equation}
and $\mathcal{M}_j$ is defined in Eq.~\eqref{eq:mjOperator}. As a further simplification, we assume that the temperature of the buffer gas is negligible, and that each of the three components of the mean energy before the collision are approximately equal in magnitude, i.e., $\langle E_x \rangle = \langle E_y \rangle = \langle E_z \rangle = \frac{1}{3}\langle E \rangle$. The ratio of the post-collision energy, $\langle E' \rangle = \sum_j \langle E'_j \rangle$, to the pre-collision energy $\langle E \rangle$ is then given by,
\begin{equation}
\frac{\langle E' \rangle}{\langle E \rangle} = \frac{1}{ (1+ \tilde{m})^2}   + \frac{ \tilde{m}^2}{9 (1+\tilde{m})^2 } \sum_{j,k\in (x,y,z)} \frac{   c_{0,j}^2 W_k^2 \beta_j^2}{ c_{0,k}^2 W_j^2 \beta_k^2   } \mathcal{M}_j[ (\mathrm{ce}_k(\tau)^2  + \mathrm{se}_k(\tau)^2 )^{-1}].
\end{equation}
For non-zero temperatures, we may solve Eq.~\eqref{eq:centralCollisionEnergy} for the steady-state values of $\langle E_j \rangle$ by setting $\langle E'_j \rangle = \langle E_j \rangle$, and solving the resulting set of linear equations. For sufficiently large values of $\tilde{m}$, no physically meaningful solution exists corresponding to values of $\tilde{m}$ greater than the critical mass ratio.  

\section{Secular phase distribution} \label{eq:appendix:secularPhaseDist}
The probability that a collision takes place at a given location $\mathbf{r}$ in a time interval $\Delta t$ is proportional to the density $\rho(\mathbf{r})$ of the buffer gas,
\begin{equation} \label{eq:collisionGivenPosition}
p(c | \mathbf{r})= k_c \Delta t \rho(\mathbf{r}) .
\end{equation}
where the notation $c | \mathbf{r}$ indicates the probability of a collision ($c$) at a specific position $\mathbf{r}$ and where $k_c$ is the collision rate constant. By employing Bayes' theorem, this may be converted to the probability for the ion to be at position $\mathbf{r}$ at the time of a collision \cite{riley10a},
\begin{equation} \label{eq:positionConditionedCollision}
p(\mathbf{r}| c) = \frac{p(c| \mathbf{r}) p(\mathbf{r})}{\int p(c| \mathbf{r}) p(\mathbf{r}) d \mathbf{r}}.
\end{equation}
To proceed, we make the simplification that the density of the buffer gas changes sufficiently slowly such that it depends only on the secular position of the ion, and take the secular position to be given by $r_j = A_j \cos (\phi_j + \omega_j t) = A_j \cos \tilde{\phi_j}$. This approximation is appropriate for a buffer gas which is not strongly localised, i.e.,  the density of the buffer gas does not vary significantly over the length scale given by the amplitude of the micromotion, that is, the density is approximately constant over an interval of width $r_j q_j$ centred at $r_j$.  For a given value of $A_j$, the probability for a component of the secular position to take a specific value in the interval $[-A_j,A_j]$ is \cite{robinett95a},
\begin{equation} \label{eq:secularPositionComponentProb}
p(r_j) =  \left(\pi \sqrt{A_j^2 - r_j^2}\right)^{-1},
\end{equation}
and so, assuming that the position for each axis is independent,
\begin{equation}\label{eq:secularPositionProb}
p(\mathbf{r}) =   \prod_{j \in (x,y,z)} p(r_j) = \prod_{j \in (x,y,z)}  \left(\pi \sqrt{A_j^2 - r_j^2}\right)^{-1}.
\end{equation}
Typically, the neutral buffer gas is confined in a potential which is approximately harmonic at the centre of the trap, such that the density of the buffer gas follows a Gaussian density distribution,
\begin{equation} \label{eq:bufferGasGaussianAppendix}
\rho(\mathbf{r})=\rho_x(r_x) \rho_y(r_y) \rho_z(r_z),
\end{equation}
where,
\begin{equation}
\rho_j(r_j) =\frac{1}{\sqrt{2 \pi } \sigma_j} e^{-\frac {r_j^2}{2 \sigma_j^2} }.
\end{equation}
Substituting Eqs.~\eqref{eq:collisionGivenPosition}, \eqref{eq:secularPositionProb} and \eqref{eq:bufferGasGaussian} into Eq.~\eqref{eq:positionConditionedCollision} and evaluating the integral produces,
\begin{equation}
p(\mathbf{r}| c) = \prod_{j \in(x,y,z)} \frac{\exp \left(\frac{1}{2} \left(\frac{A_j^2-2 r_j^2}{2\sigma_j^2}\right)\right)}{\pi \sqrt{A_j^2-r_j^2} I_0\left(\frac{A_j^2}{4 \sigma_j^2}\right) },
\end{equation}
where $I_n(x)$ is the modified Bessel function of the first kind \cite{olver2010a}. Employing a change of variables $r_j = A_j \cos \tilde{\phi_j}$, we obtain the distribution for the instantaneous secular phase for the motion along each axis at the time of a collision,
\begin{equation} \label{eq:phiDistLocalisedAppendix}
f_{\tilde{\phi_j}}(\tilde{\phi_j} | c)=\frac{1}{2\pi}\frac{e^{-\frac{A_j^2 \cos (2 \tilde{\phi_j})}{4 \sigma_j^2}}}{  I_0\left(\frac{A_j^2}{4 \sigma_j^2}\right) } .
\end{equation}
 
 \section{Analytical expression for \texorpdfstring{$\langle \eta_1 \rangle$}{mean value of n1} } \label{section:eta1Expression}
 In the main text, it is demonstrated that in the presence of a buffer gas confined in a harmonic potential, the change in the secular energy of an ion as the result of a collision is approximated by,
\begin{equation}
E' \approx (\eta_0 - \eta_1 E)E + \epsilon.
\end{equation}
Here, we provide an expression for the expectation value of $\eta_1$, i.e., the value averaged over all the variables contributing to the outcome of a collision. This value is obtained by substituting $\phi_j = \tilde{\phi_j} - \beta_j \tau$ into the expression for $\eta$ as given in the Supplementary Material of Ref.~\cite{rouse18a}, then averaging the result over the distributions for $\tilde{\phi_j}$ derived in Appendix~\ref{eq:appendix:secularPhaseDist}. Applying Eq.~\eqref{eq:besselRatioExpansion} to this expression produces a set of terms independent of the secular energy, i.e., $\eta_0$, and a set of terms linearly proportional to the energy, i.e., $\eta_1$. Averaging the terms contributing to $\eta_1$ over the remaining variables produces,
\begin{equation}
\langle \eta_1 \rangle = \int \int \int \frac{f_\tau(\tau) f_{\theta_\rho}(\theta_\rho) f_{\phi_\rho}(\phi_\rho) }{\left(\tilde{m}+1\right)^2 k_B T_b}  \sum_{j \in (x,y,z)} \frac{F_j(\tau) P_j^2 \omega _{j,b}^2}{\text{c}_{0,j}^2 W_j^2 \omega _j^2}d \tau  d \theta_\rho d \phi_\rho,
\end{equation}
where the remaining average over $\tau$ is left unevaluated due to the complexity of integrals involving the Mathieu functions, and the averages over $\phi_\rho,\theta_\rho$ are left unevaluated due to the lack of an accurate analytical form for the distributions of these two variables.  In the above, $P_j$ is defined as in the main text and is a function of $\theta_\rho,\phi_\rho$ (see text following Eq.~(12))  and the function $F_x(\tau)$ is defined by,
\begin{equation} \label{eq:eta1ValLong}
\begin{split}
F_x&(\tau)  = \\&\frac{\tilde{m}^3 c_{0,y}^2 W_x^2 \beta _y^2}{24  c_{0,x}^2 W_y^2 \beta _x^2}\bigg[\text{cs}_y(\tau ) \left[2 \dot{\text{ce}}_x(\tau ) \dot{\text{se}}_x(\tau ) \sin \left(2 \tau  \beta _x\right)+\left(\dot{\text{ce}}_x(\tau ){}^2-\dot{\text{se}}_x(\tau ){}^2\right) \cos \left(2 \tau  \beta _x\right)\right]\bigg]
\\&+\frac{\tilde{m}^3 c_{0,z}^2 W_x^2 \beta _z^2}{24  c_{0,x}^2 W_z^2 \beta _x^2} \bigg[\text{cs}_z(\tau ) \left[2 \dot{\text{ce}}_x(\tau ) \dot{\text{se}}_x(\tau ) \sin \left(2 \tau  \beta _x\right)+\left(\dot{\text{ce}}_x(\tau ){}^2-\dot{\text{se}}_x(\tau ){}^2\right) \cos \left(2 \tau  \beta _x\right)\right]\bigg]
\\&+\frac{\tilde{m}^3}{12}\bigg[\text{cs}_x(\tau ) \dot{\text{ce}}_x(\tau )  \dot{\text{se}}_x(\tau ) \sin \left(2 \tau  \beta _x\right)\bigg]
\\&+\frac{\tilde{m}^2}{4}  W_x \bigg[\left(\text{se}_x(\tau ) \dot{\text{se}}_x(\tau )-\text{ce}_x(\tau ) \dot{\text{ce}}_x(\tau )\right) \sin \left(2 \tau  \beta _x\right) \\ &~~~~~~~~~~~+\left(\dot{\text{ce}}_x(\tau ) \text{se}_x(\tau )+\text{ce}_x(\tau ) \dot{\text{se}}_x(\tau )\right) \cos \left(2 \tau  \beta _x\right)\bigg]
\\&+\frac{\tilde{m}^3}{12}\bigg[3 \text{ce}_x(\tau ) \text{se}_x(\tau ) \left(\dot{\text{ce}}_x(\tau ){}^2+\dot{\text{se}}_x(\tau ){}^2\right) \sin \left(2 \tau  \beta _x\right) \\ &~~~~~+\left(\dot{\text{ce}}_x(\tau ){}^2 \left(2 \text{ce}_x(\tau ){}^2-\text{se}_x(\tau ){}^2\right)+\dot{\text{se}}_x(\tau ){}^2 \left(\text{ce}_x(\tau ){}^2-2 \text{se}_x(\tau ){}^2\right)\right) \cos \left(2 \tau  \beta _x\right)\bigg]
\end{split}
\end{equation}
where $\mathrm{cs}_j(\tau) = \matcj^2 + \matsj^2$. The functions $F_y(\tau)$ and $F_z(\tau)$ have the same general form and are found by switching a pair of indices, e.g, $F_y(\tau)$ is found by replacing $x$ with $y$ and vice versa.

 \section{Moments of superstatistical distributions} \label{section:superstatMoments}
For a general energy distribution which can be expressed as a superposition of thermal states, i.e.,
\begin{equation}
f_E(E) = \int  \frac{E^k}{(k_B T)^{k+1} \Gamma(k+1)} f_T(T) e^{-\frac{E}{k_B T}} dT,
\end{equation}
the moments are given by,
\begin{equation}
\langle E^n \rangle = \int E^n f_E(E) dE = \int \int E^n \frac{E^k}{(k_B T)^{k+1} \Gamma(k+1)} f_T(T) e^{-\frac{E}{k_B T}} dT dE.
\end{equation}
Exchanging the order of integration to first integrate over $E$ produces,
\begin{equation}
\langle E^n \rangle =   k_B^n  \frac{\Gamma(k+n+1)}{\Gamma(k+1)}\int  f_T(T) T^n dT ,  
\end{equation}
for $k + n > -1$ and where the terms independent of $T$ have been moved outside the integral. The integral itself is the definition of the expectation value of $T^n$, i.e., $\langle T^n \rangle$ \cite{riley10a}. Thus, 
\begin{equation} \label{eq:temperatureMomentToEnergyMoment}
 \langle E^n \rangle = k_B^n \frac{\Gamma(k+n+1)}{\Gamma(k+1)} \langle T^n \rangle .
\end{equation}

\section{The Bessel-Tsallis distribution} \label{section:besselTsallisDerivation}
When the buffer gas follows a Gaussian density distribution, the change in the mean energy with each collision can be approximated by,
\begin{equation}
\langle E' \rangle = \langle \eta_0 \rangle \langle E \rangle - \langle \eta_1 \rangle \langle E^2 \rangle + \langle \epsilon \rangle,
\end{equation} 
and using Eq.~\eqref{eq:temperatureMomentToEnergyMoment} we obtain,
\begin{equation}
\langle T' \rangle = \langle \eta_0 \rangle \langle T \rangle - 4 k_B \langle \eta_1 \rangle \langle T^2 \rangle + \frac{\langle \epsilon \rangle}{3 k_B}.
\end{equation}
We assume that the multiplicative term $\eta$ is the most significant source of noise, and so in the recurrence relation of the random variable $T$ the other variables can be treated as constants. We have shown previously that multiplication of the energy by a random value is equivalent to multiplying the temperature by the same random value, that is, $E' = \eta E$ is equivalent to $T' = \eta T$ \cite{rouse17a}. Neglecting the fluctuations in both the additive term and $\eta_1$, a suitable recurrence relation for the random variable $T$ is,
\begin{equation} \label{eq:temperatureRecurrence}
T' = \eta_0 T - 4 k_B \langle \eta_1 \rangle T^2 + \kappa T_b,
\end{equation}
where $\kappa T_b = \langle \epsilon \rangle / (3 k_B)$ and $\kappa$ is defined as in Eq.~\eqref{eq:kappaDef}. We solve this recurrence relation using the method in Ref.~\cite{sornette97a} by converting it to a Langevin equation for the variable $x = \ln T$.
We first consider the case where $\langle \eta_1 \rangle = 0,  \kappa T_b = 0$ to establish a suitable representation for the noise term $\eta_0$. Since $T' = \eta_0 T$, it follows that,
\begin{equation}
\ln T' = \ln \eta_0 + \ln T.
\end{equation}
We approximate the finite difference by a differential, $\frac{dx}{dt} = \ln T' - \ln T$, and separate $\ln \eta_0$ into its mean value $\mu = \langle \ln \eta_0 \rangle$ and a fluctuating term, $\hat{\zeta}(t)$, such that,
\begin{equation} \label{eq:langevinEqMultiOnly}
\frac{dx}{dt} = \mu + \hat{\zeta}(t) .
\end{equation}
This has converted the multiplicative stochastic process in terms of $T$ and $\eta$ into an additive stochastic process in terms of $x$ and $\ln \eta$. By itself, Eq.~\eqref{eq:langevinEqMultiOnly} does not produce a stable steady-state distribution for $x$ \cite{sornette97a}, and it does not include the effects of the temperature of the buffer gas or the reduction in $\eta$ due to localisation. To find a representation for these terms, we use a different approximation for the derivative, $\frac{dx}{dt} \approx \frac{T' - T}{T}$ following Ref.~\cite{sornette97a}. Using this approximation with Eq.~\eqref{eq:temperatureRecurrence} produces,
\begin{equation} \label{eq:langevinEqAllApprox}
\frac{d x}{d t} \approx \langle \eta_0 \rangle + \hat{\eta}(t) - 1 - 4 k_B \langle \eta_1 \rangle e^x  + \kappa T_b e^{-x},
\end{equation}
where we have separated $\eta_0$ into its mean $\langle \eta_0 \rangle$ and a fluctuating term $\hat{\eta}(t)$ as before. Notice that since we have used a less accurate approximation for $\frac{dx}{dt}$, this equation is defined in terms of $\eta_0 - 1$ rather than $\ln \eta_0$. However, $\langle \eta_0 \rangle -1 \approx \mu$ and the variances of $\hat{\eta}(t)$ and $\hat{\zeta}(t)$ are approximately equal \cite{sornette97a}, such that Eq.~\eqref{eq:langevinEqAllApprox} is approximately equivalent to Eq.~\eqref{eq:langevinEqMultiOnly} in the limit where $T_b = 0$ and $\langle \eta_1 \rangle = 0$. Moreover, this approximation provides a representation for the effects of both a non-zero buffer gas temperature and the non-uniform density of the buffer gas. We therefore augment Eq.~\eqref{eq:langevinEqMultiOnly} with the terms proportional to $T_b$ and $\langle \eta_1 \rangle$ of Eq. \eqref{eq:langevinEqAllApprox} to produce,
\begin{equation}\label{eq:langevinEqTemperatureLocal2}
\frac{d x}{d t} = \mu + \hat{\zeta}(t) + \kappa T_b e^{-x} - 4 \langle \eta_1 \rangle k_B e^x .
\end{equation}
We make the approximation that $\hat{\zeta}(t)$ can be modelled as following a Gaussian distribution, i.e., it represents the fluctuations in $x$ averaged over multiple collisions  \cite{honerkamp02a}. This follows from the fact that the fluctuations in $x$ are additive, and the sum of independent random variables approaches a Gaussian distribution by the central limit theorem \cite{honerkamp02a}. This approximation enables the derivation of an analytically tractable Fokker-Planck equation for the probability distribution $f_x(x)$ \cite{sornette97a},
\begin{equation} \label{eq:thermalFokkerPlanckLocal}
\frac{\sigma^2}{2} \frac{d^2}{d x^2} f_x(x) - \frac{d}{dx}\left[   (     \mu + \kappa T_b e^{-x}  - 4 \langle \eta_1 \rangle k_B e^x  ) f_x(x) \right] = 0,
\end{equation}
where $\sigma^2$ is the variance of $\hat{\zeta}(t)$. If $\langle \eta_1 \rangle = 0$, then this equation reduces to the one obtained in Ref.~\cite{rouse17a}, and a steady-state solution exists if $\mu < 0$ and $T_b \neq 0$. Conversely, if $T_b = 0$, then a solution exists only if $\langle \eta_1 \rangle$ is non-zero and $\mu > 0$. These conditions correspond to the overall drift of $x$ towards a lower or an upper bound \cite{sornette97a}. We proceed assuming that both $T_b,\langle \eta_1 \rangle$ are non-zero, such that both upper and lower bounds exist, and the existence of a steady-state does not depend on the sign of $\mu$. Subject to the boundary conditions that $f_x(x)\rightarrow 0$ for $x\rightarrow \pm \infty$, the steady-state distribution for $T$ is,
\begin{equation} \label{eq:gigTemperature}
f_T^{(L)}(T) = \frac{2^{\nu -1} \left(\frac{b}{\nu }\right)^{-\frac{\nu }{2}} T^{-\nu -1} \left(\frac{k_B^2}{E_\ell}\right){}^{-\frac{\nu }{2}} e^{-\frac{\nu }{b k_B T}-\frac{k_B T}{4 E_\ell}}}{K_{\nu }\left(\sqrt{\frac{\nu }{b E_\ell}}\right)} ,
\end{equation}
where $K_y(z)$ is the modified Bessel function of the second kind with order $y$ and argument $z$, and the superscript $(L)$ is used to indicate that this is the distribution obtained in the presence of a localised buffer gas. The parameters are defined in terms of the coefficients of Eq.~\eqref{eq:thermalFokkerPlanckLocal} as,
\begin{equation}
b = \frac{-\mu}{k_B \kappa T_b},
\end{equation}
\begin{equation}
\nu = \frac{-2 \mu}{\sigma^2} ,
\end{equation}
\begin{equation}
E_\ell = \frac{\sigma^2}{32 \langle \eta_1 \rangle},
\end{equation}
 and the distribution is normalisable if $E_\ell > 0$ and $b/\nu > 0$. These definitions for the parameters are accurate in the limit in which $\hat{\zeta}(t)$ can be approximated as following a Gaussian distribution, but do not take into account corrections due to the exact form of the distribution of $\ln \eta_0$ \cite{sornette97a,rouse17a}.

The energy distribution is defined as a superposition of thermal distributions,
\begin{equation} \label{eq:superstatsEnergyAgain}
f_E(E) = \int f_E(E|T) = \frac{E^k}{ (k_B T)^{k+1}  \Gamma(k+1) } e^{- \frac{E}{k_B T}} f_T(T)   d T.
\end{equation}
Evaluating Eq.~\eqref{eq:superstatsEnergyAgain} using Eq.~\eqref{eq:gigTemperature} produces,
\begin{equation} \label{eq:besselTsallisDistK}
f_E^{(BT)}(E) =  \frac{ E^k \left(\frac{b E}{\nu }+1\right)^{-\frac{1}{2} (k+\nu +1)} \left(\frac{b}{\nu  E_\ell}\right){}^{\frac{k+1}{2}} K_{k+\nu +1}\left(\sqrt{\frac{E}{E_\ell}+\frac{\nu }{b E_\ell}}\right)}{2^{k+1} \Gamma (k+1) K_{\nu }\left(\sqrt{\frac{\nu }{b E_\ell}}\right)}.
\end{equation}
The moments of this distribution are difficult to evaluate directly due to the complexity of integrals involving the Bessel function. However, the moments of Eq.~\eqref{eq:gigTemperature} may be easily calculated by evaluating $\int T^n f_T(T) dT$, and applying Eq.~\eqref{eq:temperatureMomentToEnergyMoment} produces,
\begin{equation}
 \langle E^n \rangle = \frac{2^n \Gamma (k+n+1) \left(\frac{b}{\nu  E_\ell}\right){}^{-\frac{n}{2}} K_{\nu -n}\left(\sqrt{\frac{\nu }{bE_\ell}}  \right)}{\Gamma (k+1) K_{\nu }\left(\sqrt{\frac{\nu }{bE_\ell}}    \right)}.
\end{equation}
The mean energy $\langle E \rangle$ evaluated using this expression is defined as long as $E_\ell > 0$ and $b/\nu > 0$.  
\section{Parameter estimation} \label{section:parameterEstimation}
Although the values of $\mu,\sigma,\langle \eta_1 \rangle$ required to calculate $\nu,b,E_\ell$ are in theory defined in terms of the mass ratio and trapping parameters, in practice these cannot be accurately evaluated a priori due to the fact that the distributions of $\ln \eta_0$, $\phi_\rho$ and $\theta_\rho$ are not known analytically. Instead, the required values can be obtained from numerical simulations by sampling the distribution of $\eta$. To do so, a series of collisions is simulated to produce a value of $E$ under the same conditions as for the simulations used to obtain the energy distribution. For the final collision, the buffer gas temperature is set to $T_b = 0~$K, such that the change in energy is given by $E' = \eta E$. Dividing the post-collision energy by the pre-collision energy provides a value for $\eta$, and repeating this process (typically for 1'000'000 iterations) produces a set of values of $E,\eta$. From these, the coefficients for the linear expansion $\eta = \eta_0 - \eta_1 E$ are obtained by least-squares linear regression, and we take $\langle \eta_1 \rangle$ to be equal to the value of $\eta_1$ extracted by this method. If the density of the buffer gas is set to a uniform distribution for the final collision in addition to setting the temperature equal to zero, the $\eta_1$ term is eliminated, and we have $\eta = \eta_0$. The values of $\mu = \langle \ln \eta_0 \rangle$ and $\sigma^2 = \langle (\ln \eta_0)^2 \rangle - \mu^2$ are then calculated from the values of $\eta_0$ obtained in this manner. This method is in general found to produce acceptable results except for the particular case of $\tilde{m} = 1, \omega_{r,b} = 100~$Hz, for which the ion's energy in the steady-state remains small enough that $\eta_1$ cannot be accurately estimated from collisions. Consequently, for that mass ratio $\eta_1$ is obtained by setting the initial temperature of the ion to $T = 1~$K and performing a single collision, rather than allowing the ion to reach the steady-state first. This leads to a slightly different value of $\eta_1$ than would be obtained in the steady-state, but in practice for this mass ratio the energy of the ion in the steady-state remains sufficiently low that the effect of localisation is unimportant, i.e., $E << E_\ell$, and so the error in $\eta_1$ does not significantly affect the shape of the distribution. Indeed, the steady-state energy distribution is found to essentially follow Tsallis statistics over the range of energies sampled in numerical simulations.

Alternatively, the distribution may be fit to numerical data through maximum-likelihood estimation. Analytical expressions for the maximum-likelihood estimates of the parameters $b,\nu,E_\ell$ have not yet been obtained due to the complexity of derivatives of the Bessel function with respect to $\nu$.  Thus, the estimation is performed numerically with respect to the parameters $\tilde{b} = b/\nu, \nu, E_\ell$. The use of $\tilde{b}$ ensures that this parameter is strictly positive, reducing the range of values to optimise over and eliminating the constraint that $b$ must have the same sign as $\nu$. The parameters of the exponential-Tsallis distribution and the standard Tsallis distribution are also found through numerical maximum likelihood estimation.

\end{document}